\documentclass[twocolumn]{revtex4-2}
\usepackage[utf8x]{inputenc}
\usepackage[english]{babel}

\usepackage{graphicx}
\usepackage{xcolor}
\usepackage{xfrac}
\usepackage{amsmath}
\usepackage{float}
\usepackage{lipsum}
\usepackage{hyperref}

\usepackage[all]{hypcap} 

\usepackage[capitalize]{cleveref}
\usepackage[normalem]{ulem}

\graphicspath{{./}}

\begin{document}
	
	\title{Domain convexification: a simple model for invasion processes}
	\author{David \textsc{Martin-Calle}; Olivier \textsc{Pierre-Louis}}
	\date{\today}
	
	\begin{abstract}
		We propose an invasion model where domains grow up to their convex hulls and merge when they overlap. 
        This model can be seen as a continuum and isotropic counterpart of bootstrap percolation models. 
        From numerical investigations of the model starting with randomly scattered discs in two dimensions, we find an invasion transition that occurs via macroscopic avalanches.
		The disc concentration threshold and the sharpness of the transition are found to decrease as the system size is increased. 
        Our results are consistent with a vanishing threshold in the limit of infinitely large system sizes. However this limit could not be investigated by simulations. 
        For finite initial concentrations of discs, the cluster size distribution presents a power-law tail characterized by an exponent that varies approximately linearly with the initial concentration of discs. 
        These results at finite initial concentration open novel directions for the understanding
        of the transition in systems of finite size.
        Furthermore, we find that the domain area distribution has oscillations with discontinuities.
        In addition, the deviation from circularity of large domains is constant.
		Finally, we compare our results to experimental observations on de-adhesion of graphene induced by the intercalation of nanoparticles.
	\end{abstract}
	
	\maketitle
%

	\section{Introduction}

\begin{figure*}
	\centering
	\includegraphics[width=0.8\linewidth]{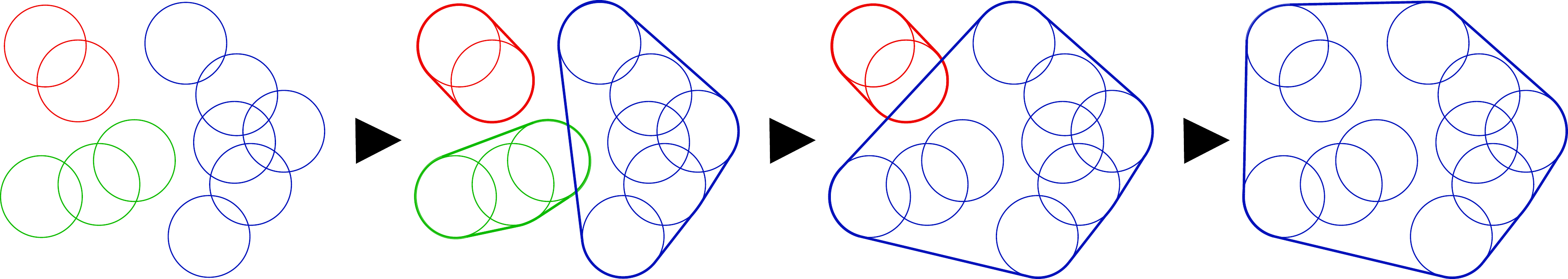}
	\caption{Schematic of the iterative convexification and merging process. 
	The process is repeated as long as merging is possible. We analyze the
	final state at the end of this process.}
	\label{fig:illust_convex}
\end{figure*}

Invasion transitions, where domains invade the entire space, 
are found in many two-dimensional systems~\cite{Livi2017,Krapivsky2010}
ranging from phase transitions in magnets to de-adhesion transitions
and spreading of species in ecology models. 
In this paper, we propose a simple two-dimensional model 
for isotropic invasion in continuous space where each domain grows up
to its convex hull, thereby invading the concave parts
along its periphery. 
If this growth process up to the convex hull, hereafter denoted as
convexification, leads to overlapping with other domains,
then these domains are merged.
Merging of domains produces new 
concavities, leading to additional domain growth.
A schematic of the model starting with randomly
scattered discs on the plane is shown in \cref{fig:illust_convex}.
Our numerical investigation of this model suggests that 
it exhibits a invasion transition as the initial density
of discs is increased.

Our model constitutes a continuous and isotropic 
counterpart of the bootstrap percolation model~\cite{Aizenman1988,Adler1991,DeGregorio2021}. 
Bootstrap percolation models are on-lattice models
where domains grow irreversibly at sites
that have a sufficient number of bonds with the
existing domains. For a large enough threshold in the number of bonds ($2$ for the square
lattice), and starting from random initial conditions with a 
given initial density of occupied sites, the 
domains grow up to a final state which is either
a set of disconnected domains with a finite domain density, or a single domain
that has invaded the whole system.
A transition between these two regimes 
occurs at a finite initial density of occupied sites for systems with a finite size.
However, this finite initial density threshold 
is a finite-size effect, and in the limit of very large
systems, the threshold of the transition occurs at
vanishing initial densities~\cite{Holroyd2003}.
The convergence to the limit of large system size in bootstrap percolation
is notoriously slow~\cite{DeGregorio2005}. Our simulations show that 
the transition threshold of the convexification model decreases with
increasing system size. 
Our numerical evidences are insufficient
to conclude that the threshold vanishes or not for very large sizes.
However, in the range of system sizes that we have explored,
we find that the threshold and the width of the transition
decrease when the system size increase.

Moreover, the bootstrap percolation transition 
is known to occur in a discontinuous way via a macroscopic avalanche when 
increasing the initial density of occupied sites~\cite{Manna1998,Farrow2005,Farrow2007}.
Such a discontinuity has attracted much attention
in the literature. Other percolation
models were also found to exhibit strongly discontinuous 
transitions, often called explosive percolation~\cite{DSouza2015,DSouza2019}.
Macroscopic avalanches have also been observed in the depinning
transition of the low-temperature
random field Ising model~\cite{Sethna1993,Rosinberg2003,Nandi2016}.
In both  bootstrap percolation and depinning transitions,
the discontinuous character of macroscopic avalanches
is associated with strong finite-size effects that are rooted
in rare events~\cite{DeGregorio2005,Nandi2016}.
For large-enough systems, we also observe that
the transition takes the form of a macroscopic avalanche that
can be triggered by a microscopic perturbation,
for example, the addition of a single disc to the initial condition.

The bootstrap percolation and depinning
transitions occur in simple paradigmatic models that are expected
to catch the essence of a wide variety of physical phenomena,
from the jamming transition to crack propagation.
In like manner, we expect our model
to pertain to a large class of systems where
invasion transitions can be observed,
the main novelty of our model being its isotropic lattice-free character.

A first class of phenomena 
that may be described by the convexification model
is line-tension induced growth of two-dimensional domains
with random impurities or defects. 
In contrast to
the random field Ising model, the randomness then only
originates in the initial condition:
we assume that one type of magnetization 
is imposed in some randomly-placed disc-shaped zones.
Outside these zones, the interface between the 
two types of magnetization moves freely, and 
is subject to standard motion by curvature
driven by line tension~\cite{Livi2017,Krapivsky2010}.
Hence, these interfaces straighten and build
the convex hull of the initial zones.
Due to the equivalence between lattice gas models
and the Ising model~\cite{Saito1996}, this process also describes
the dynamics of adsorption of a monolayer of molecules (or particles)
with short-range attractive interactions 
on a flat substrate in the presence of localized defects or impurities which 
enforce the initial coverage of some circular zones.
Another similar system is the imbibition
in a Hele-Shaw geometry~\cite{Paterson1995,Horvath1991,Rubio1989} with strongly-wetting defects,
a geometry that mimics imbibition of porous media.

A second class of physical phenomena
that could be described
by domain convexification is third-body induced
de-adhesion transitions. These transitions
are relevant to friction and its coupling with wear,
but also to intercalation of particles, molecules, or atoms inside layered
materials such as graphite~\cite{LiiRosales2020,Zong2010,Wang2016}.
A simple view of the de-adhesion transition is the
following. When particles are intercalated
in the contact zone between two bodies,
each particle leads to a local de-adhesion zone around it.
When these zones overlap, the concave parts of their periphery can be rounded, 
leading to a growth of the de-adhesion zone. This growth can lead to
new overlap with other de-adhesion zones, and the growth of de-adhesion zones occurs again.
When the density of particles is large enough, this leads to de-adhesion
of the full contact zone.
This generic phenomena has been observed
in the case of de-adhesion of graphene
with intercalated nanoparticles~\cite{Yamamoto2012}.
One advantage of using a two-dimensional material
like graphene is the possibility of imaging
the detailed geometry of detachment zones induced
by nanoparticles. In these experiments,
a de-adhesion transition is observed when
density of intercalated nanoparticles, or the rigidity
of the membrane are increased. 
The convexification model appears as a
idealization of this process which assumes that
detachment zones simply grow up
to the convex hull of the detachment zones
that would be induced by each particle independently.
The physics of de-adhesion is expected 
to be more complex than this simple model, since the de-adhesion zones
might grow only partially towards their convex hull, or grow to a different shape.
Also, other features are expected depending on the precise adhesion properties and
elastic properties of the system, such as wrinkles
and conical singularities~\cite{Yamamoto2012,Guedda2016,Zhang2014,Pereira2010}.
However, we hope to catch the some generic features
of this phenomenon with our simple model,
and below, we report a quantitative analysis 
of the intercalation-induced graphene de-adhesion
observed in experiments.

The convexification model could also be relevant 
for the modeling of invasion in population dynamics.
Indeed, most models for the invasion of a population in continuous space are
based on ingredients that lead to finite propagation speed
for straight fronts~\cite{Mollison1972,Kuperman1999,Abramson2003,Hanert2011}. 
Our model proposes a different paradigm,
and suggests a scenario of species invasion 
with no net growth for straight fronts. Instead,
the invading domain could grow
only via the process of convexification. A simple
mechanism for this process is straight-line travelling
of individuals between different points of the domain,
and merging of domains where individuals meet (due, e.g., to breeding
).
Since the ensemble of straight lines starting and ending
in a domain cover its convex hull, we then obtain
the convexification model. Our results
can be translated as 
the following statement for population dynamics:
above a critical initial density of randomly placed domains
where the species,
the whole system will be invaded.
Our results suggest that starting from a 
randomly scattered population
that has no propensity to spread,
invasion and merging into a single group
could be triggered by the motion of individuals in straight lines
within their own population domain.

Finally, we would like to point out 
the possible relevance of our model for 
classification and clustering
of complex sets of points or domains in the plane.
In statistics and data analysis, one of the elementary question that is at the root of many
clustering algorithms is the linear 
separability of sets of points~\cite{BenIsrael2006,Hardy1996,JoseGarcia2016}. 
In our two-dimensional plane, two sets of points or two domains of the plane are
said to be linearly separable if they
can be separated by a straight line.
Since two disjoint convex clusters can always be separated by a straight 
line, they are linearly separable from each other.
The convex hull of a set of points is actually a standard 
tool for studying linear separability~\cite{Elizondo2006}.
Moreover, the straight line that maximizes the distance
from the points 
of two clusters is called the maximum margin hyperplane,
and its distance to the closest points is called the maximum margin~\cite{Steinwart2008,Wang2005}. 
It is clear that in our model two clusters will not merge if
the maximum margin between their disc centers is
larger than the disc radius $r_d$.
As a consequence, the condition of linear separability of clusters 
with maximum margins larger than $r_d$ is equivalent to
our condition for stopping the convexification procedure.
In other words, the invasion transition corresponds
to the threshold above which it is not possible to 
separate linearly clusters of randomly scattered points 
with a maximum margin larger than $r_d$.

A closely related subject is the separation of objects
in images using their convex hulls~\cite{Jayaram2016}. As discussed in Refs.~\cite{Malladi1995,Malladi1996}
the construction of convex hulls can be obtained by 
the propagation of the boundary of the domain
with a modified motion by curvature equation governing 
the normal velocity $v_n$ 
\begin{align}
    v_n=-\min(\kappa,0)
\end{align}
where $\kappa$ is the local curvature (positive for convex domains).
This model is actually similar to the models discussed
above for the growth of magnetization models
or monolayers from initial defects.

In the following, we will start with a detailed description of the model
and its numerical analysis in \cref{s:model}.

The simulation results are reported in \cref{s:simulation_results}.
We discuss the main features of the invasion transition in \cref{s:transition},
namely, macroscopic avalanches and the decrease of the threshold and width of the transition
as the system size increases. 
In \cref{s:order_param}, we introduce an order parameter which measures the average fraction of the 
system that is not invaded.
In \cref{s:size_distribution,s:area_distribution}, we discuss the domain size and area distributions.
For finite concentrations of initial discs, the tails of the domain size distribution are found to present
a power-law behavior with a continuously varying exponent.
Moreover, the domain area distribution is found to have oscillations
accompanied with discontinuities.
We then report shortly on domain shapes in \cref{s:cluster_shape}. 
We find that the average shape of large domains presents
a constant deviation from circularity.

Further discussions and analysis of the transition
are reported in \cref{s:discussion_transition}.
After a comparison with on-lattice bootstrap percolation in \cref{s:comparison_boostrap_percolation},
we distinguish two regimes for the transition:
a first regime at small concentrations leading to
a behavior that is similar to the asymptotic regime
of bootstrap percolation discussed in \cref{s:transition_C_to_zero}, 
and a second regime at finite concentrations discussed in \cref{s:transition_C_non_zero}.

In \cref{s:experiments}, we compare our results with experiments
where the intercalation of nanoparticles leads to the de-adhesion of graphene~\cite{Yamamoto2012}. 
We find that experimental results can
be described quantitatively with our model assuming a finite size effect
involving around 200 particles.

Finally, we provide a brief summary of our results in \cref{s:conclusion}.

	\section{Model}
\label{s:model}

We consider a two-dimensional system.
The initial condition consists
in randomly
placing $N_d$ discs of radius $r_d$ in a two-dimensional
system of area $A_{syst}$~\footnote{
More precisely, the centers of the discs are chosen with a continuous 
uniform probability distribution on the system area.
}.  The concentration 
of discs is characterized by the dimensionless concentration
\begin{align}
    C=\frac{N_d\pi r_d^2}{A_{syst}}.
\end{align}
In the following, a {\it cluster} will denote one of the groups of discs
resulting from the convexification process.
In addition, {\it a domain} will denote 
the region of the plane that corresponds
to the convex hull of a cluster.

The initial domains correspond to the zones 
of the plane that are covered one disc, or several overlapping discs.
The overlapping discs belonging to one initial domain
in the plane correspond to percolation clusters of continuum disc percolation~\cite{Gawlinski1981}.
The average area fraction not covered by the disc percolation domains is~\footnote{
This relation is exact for periodic systems and asymptotically true
in finite systems of large size when the effect of boundary conditions
can be neglected.}
\begin{align}
    \Phi_p(C)={\rm e}^{-C}.
    \label{eq:coverage_discs}
\end{align}
The percolation threshold above which infinite percolation clusters
exist in large systems is $C_p\approx1.128$~\cite{Quintanilla2000}.

Starting from this random initial condition with
disc percolation clusters, we apply the following iterative convexification procedure.
First, we consider that any domain grows up to its convex hull, this is 
the convexification step.
Then, domains that overlap are merged, this is the merging step. 
We apply iteratively these two steps up to the 
situation where no new merging occurs in the merging step.
A schematic of the convexification process is shown in \cref{fig:illust_convex}.

Such an iterative algorithm aims at mimicking the invasion processes
that have been discussed in the introduction. Note
that the 
precise dynamics of the interface that describes
how the domain boundary grows up to the convex hull 
in the convexification step is not described.
However, the final state composed of a set
of disconnected convex domains is clearly unique, and this final state
is the focus of our analysis.

More precisely, we wish to investigate the dependence
of the final state on the two dimensionless numbers that govern
the problem: the disc concentration $C$, and
the normalized system size
\begin{align}
    \bar A_{syst}=\frac{A_{syst}}{\pi r_d^2}.
\end{align}
We have performed numerical simulations
where the variation of $C$ is obtained by adding discs one by one
so as to increase $N_d$ for a fixed $\bar A_{syst}$.
Once a new disc is added, we iterate convexification and
merging steps in the system, and the algorithm
converges to a new final state.
Some details on numerical techniques
are reported in \cref{a:numerics}.

Finally, we report both on systems with periodic boundary
conditions, and circular systems with fixed boundary conditions of radius 
$R_{syst}$ and area $A_{syst}=\pi R_{syst}^2$~\footnote{
In circular systems, the centers of the discs are placed randomly in
the system. Hence, when the center of a disc 
is close to the edge of the system, part of this discs may lie outside the system.}.
The investigation of periodic systems requires a careful
definition of the convexity of a domain.
This definition is discussed in \cref{a:periodic_BC}.

	\section{Results}
\label{s:simulation_results}
	
	\subsection{Transition}
\label{s:transition}

\begin{figure}
    \centering
    \includegraphics[width=\linewidth]{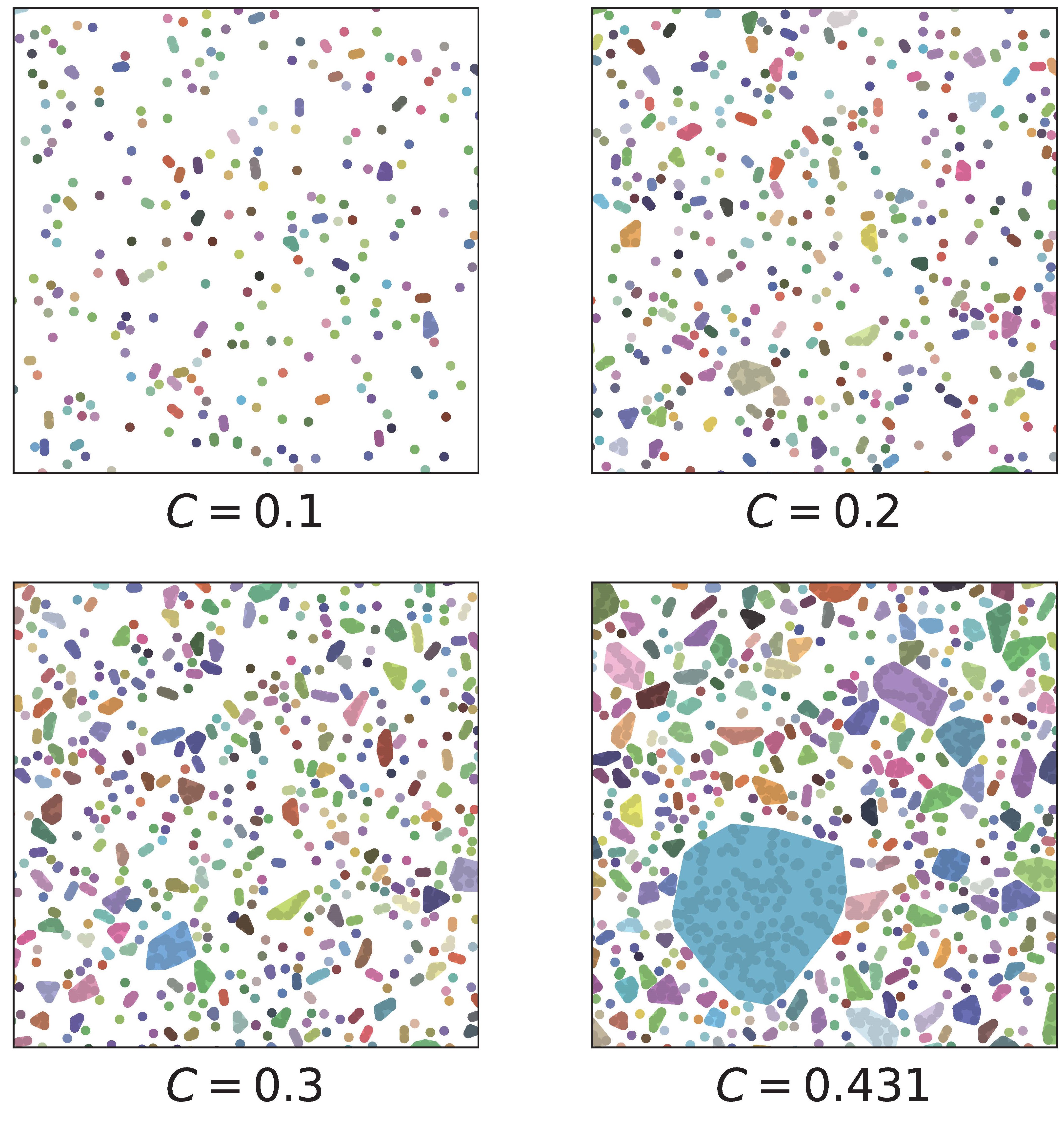}\caption{Zoom on a square region $\left(100 r_d \times 100 r_d\right)$ of the system, at various values of $C$. All domains are displayed, and discs inside them can be seen within the domains. The bottom right figure, at $C=0.431$, corresponds to the state of the system just before the transition, where the addition of a single disc triggers a macroscopic avalanche leading to the invasion of the whole system. The normalized system area is $\bar{A}_{syst}=10^5$.}\label{fig:four_steps}
\end{figure}

\begin{figure}
	\centering
	\includegraphics[width=\linewidth,trim=33 33 33 33,clip]{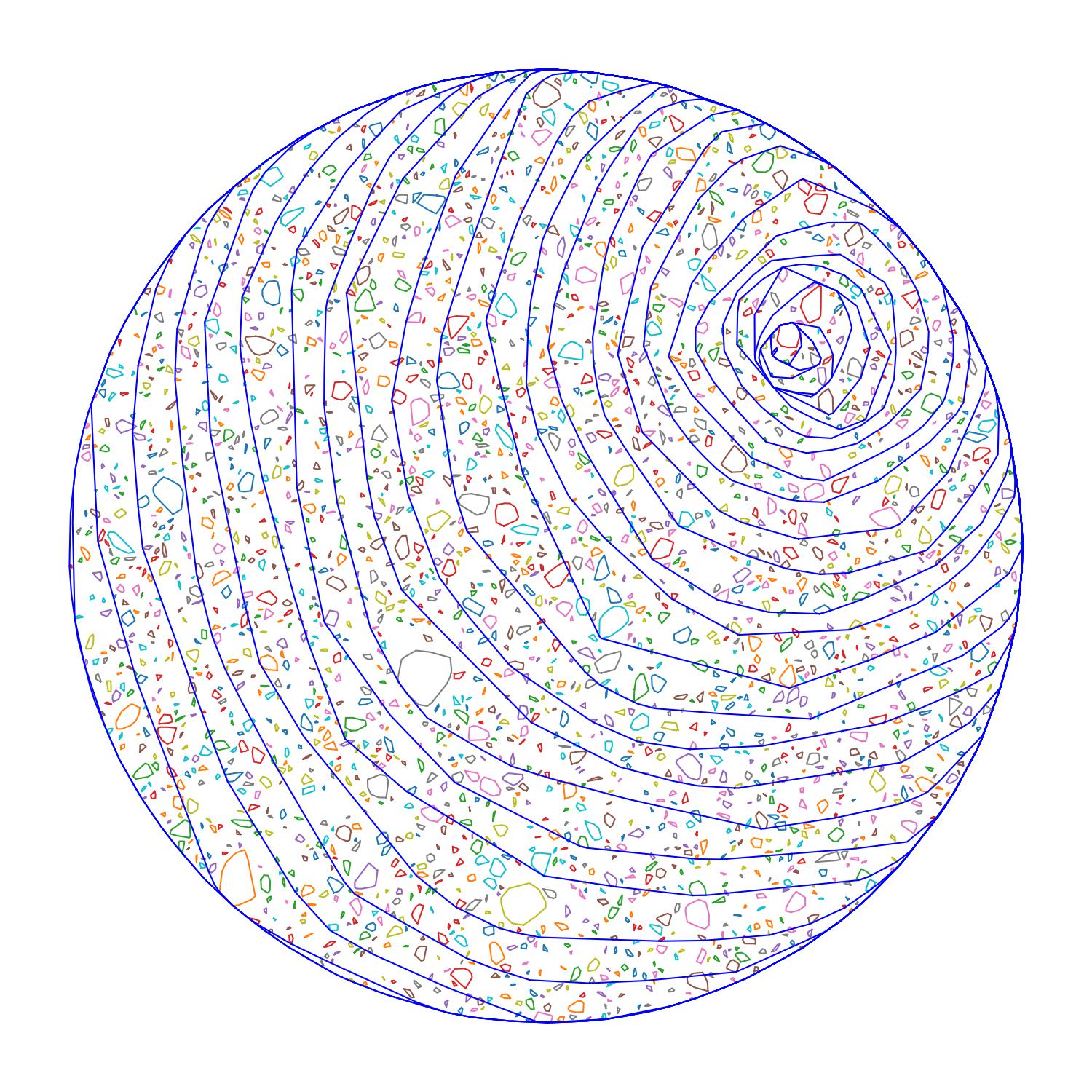}
	
    \caption{Macroscopic avalanche triggered by the addition of a single
	disc, and invading the whole system. This corresponds to the same simulation as in \cref{fig:four_steps}.
    The growing domain is shown in blue every five iterations (where one iteration 
    corresponds to the merging of the invading cluster with all the clusters it is in contact with). 
    Polygons linking disc centers are shown to indicate the domains. 
    Small clusters containing less than 8 discs in their periphery are not shown.
    $\bar{A}_{syst}=10^5$, the figure corresponds to the avalanche induced by the adding of the $43122$th disc.
    \label{fig:convex_avalanche}
    }
	
\end{figure}

A first remarkable result is obtained over one realization,
i.e., for one given sequence of added discs,
and when the system is large enough, approximately when $\bar A_{syst}\geq 10^4$.
We then observe a sharp invasion transition with the addition of a single disc.
Before this critical disc is added, the final states
are composed of a distribution of finite-size domains,
as shown in \cref{fig:four_steps}.
However, when the critical disc is added, a macroscopic
avalanche of convexification spans the whole system,
ultimately leaving only one domain that engulfs all the discs
of the system. Snapshots of the convexification 
process when adding the critical disc are reported in \cref{fig:convex_avalanche}.

\begin{figure}[h!]
	\centering
	\includegraphics[width=0.8\linewidth]{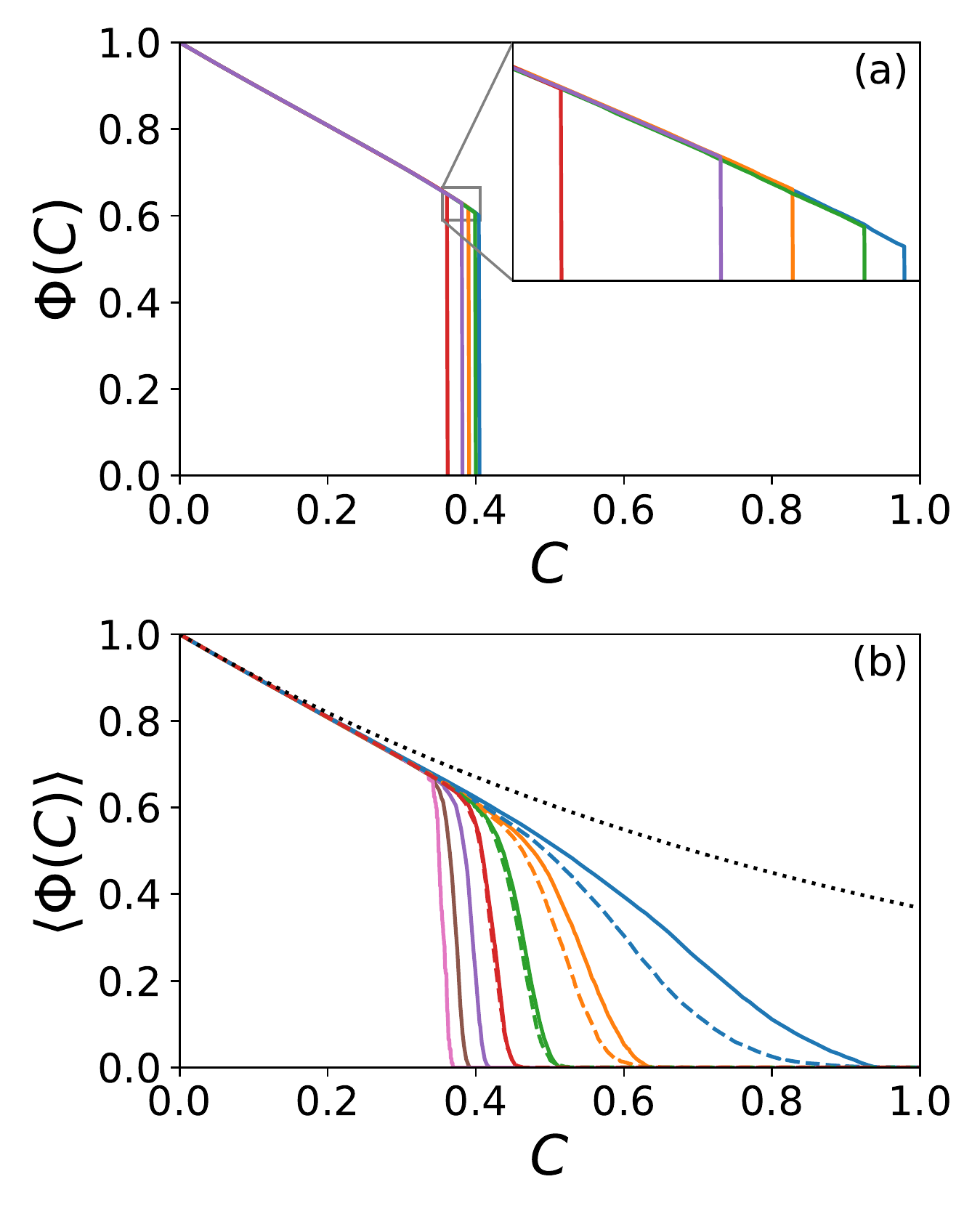}    
	\caption{$\Phi(C)$ as a function of the normalized disc concentration $C$.
	(a) $\Phi (C)$ obtained from five
	different simulations with a fixed system area  $\bar{A}_{syst}=10^6$
	and fixed boundary conditions. 
	The disc concentration $C$ is increased by adding discs one by one,
	and keeping a constant disc radius $r_d$. 
	(b) Average $\langle \Phi(C)\rangle$ for a fixed $C$. 
	Full lines: fixed boundary conditions. Dashed lines: periodic boundary conditions. 
	From right to left: 
	$\bar{A}_{syst}=[10^2,10^3,...,10^8]$ averaged over $[10^3,10^3,10^5,10^4,10^3,10^3,60]$ realizations. 
	Dotted line: order parameter without convexification $\Phi(C)=e^{-C}$. 
	}
	\label{fig:phi_C_1real}
\end{figure}

The fraction $\Phi(C)$ is defined as the ratio
of the system area that is not covered by the domains
over the system area $A_{syst}$.
In \cref{fig:phi_C_1real}(a), $\Phi(C)$ is plotted
for several independent realizations.
As expected, we have 
\begin{align}
    \Phi(C)<\Phi_p(C)
\end{align}
for any realization.
Moreover, a discontinuous transition is observed in each realization.
We observe that the value of $C$ where the transition occurs varies from
one realization to another. Macroscopic percolation clusters
are expected to be found at the percolation transition when $C\rightarrow C_p$. Since the 
convex hull of these macroscopic clusters spans the whole system,
$C_p$ should be an upper bound for the transition.
Indeed, the transitions always occurred well before the disc percolation transition
in our simulations.
In addition, the curve $\Phi(C)$ before the transition seems to be independent
of the location of the transition, i.e.,
we have observed no measurable precursors or deviations
before the appearance of the transition,
as seen in the inset of \cref{fig:phi_C_1real}(a).

\begin{figure}[h!]
		\centering
		\includegraphics[width=0.75\linewidth]{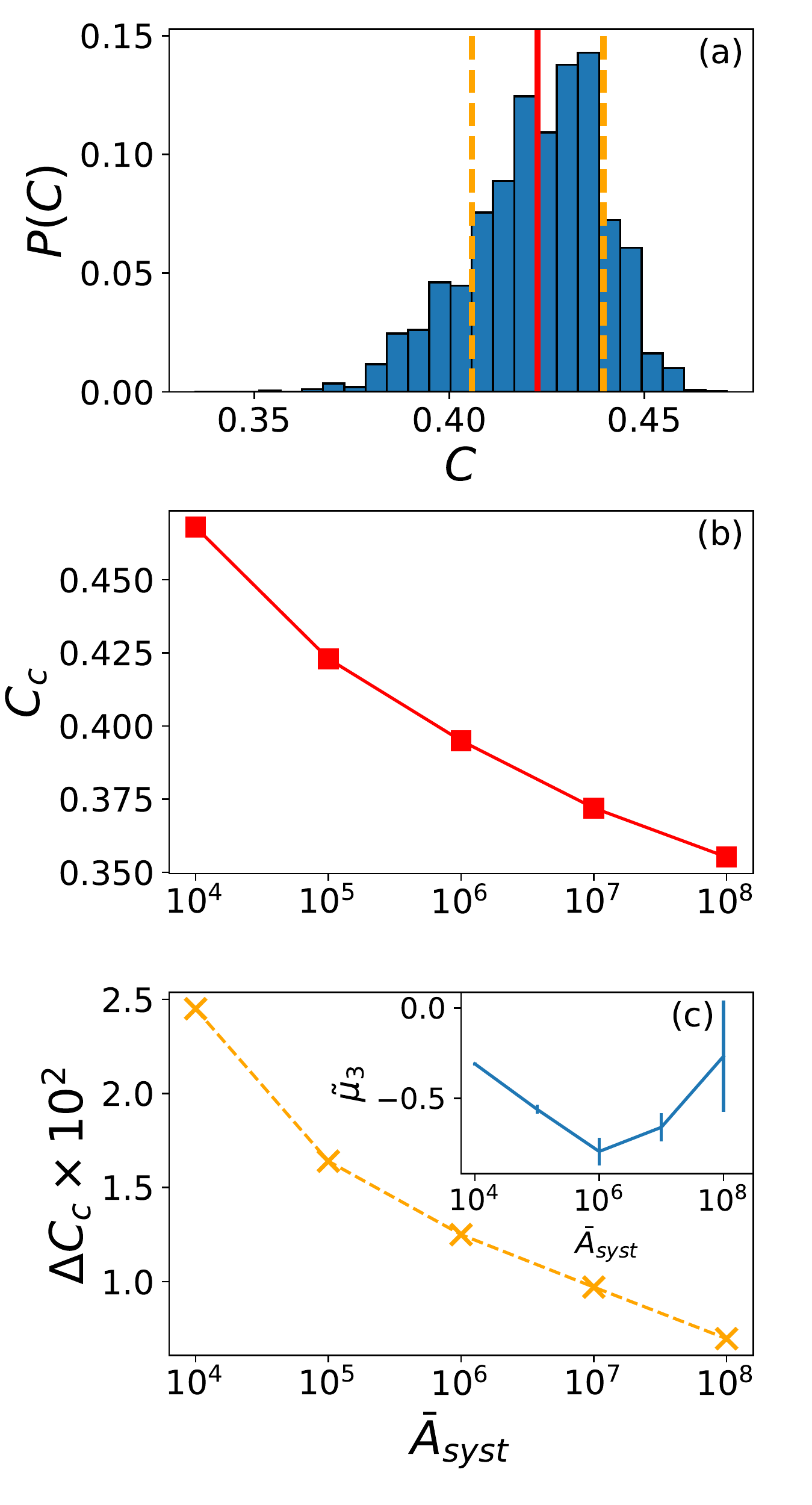}
		\caption{Transition for fixed boundary conditions.
		(a) Probability distribution of the transition point $C_c$
		from $10^4$ runs
		with $\bar{A}_{syst}=10^5$.
		Solid vertical line: average $C_c$. Dashed vertical lines: average plus or minus one standard deviation $\Delta C_c$.
		(b) Mean transition threshold ${C}_{c}$ as a function of $\bar{A}_{syst}$. 
		Averages are performed on the same number of realizations as in \cref{fig:phi_C_1real}.
        (From the points from left to right, standard errors of the mean are: [$7.743\times 10^{-5}$, $1.685\times 10^{-4}$, $3.944\times 10^{-4}$, $3.151\times 10^{-4}$, $9.1\times 10^{-4}$])
		(c) Transition width $\Delta C_{c}$ (standard deviation) as a function of $\bar{A}_{syst}$. 
        Inset: skewness of transition distributions. Errorbars correspond to plus or minus one standard error of the skewness.
		}
		\label{fig:transition_average}
	\end{figure}

An example of distribution of transition points
for $\bar{A}_{syst}=10^5$ is provided in \cref{fig:transition_average}(a).
The average $C_c$ and the standard deviation $\Delta C_c$ of this
distributions of transitions are plotted in \cref{fig:transition_average}(b,c)
as a function of $\bar{A}_{syst}$. 
They both decrease with increasing system size.

In addition, the distribution of transition points also exhibits 
a skewness that is negative for all values of $\bar{A}_{syst}$.
A plot of the skewness of the distribution is also reported in 
the inset of \cref{fig:transition_average}(c).

\subsection{Order parameter}
\label{s:order_param}

Another way to represent the distribution
of the transition is to plot the average $\langle \Phi(C)\rangle$
over many realizations, as shown in \cref{fig:phi_C_1real}(b).
The average fraction $\langle \Phi(C)\rangle$ can be used
as an order parameter to characterize percolation transitions~\cite{Quintanilla2000,Yamamoto2012,Stauffer2018}.

At small $C$, the behavior
of $\langle \Phi(C)\rangle$ is independent of the system size $\bar{A}_{syst}$.
An expansion for $C\rightarrow 0$ up to second order in $C$ can be obtained 
when considering only isolated discs and domains composed of two discs. 
Such an expansion is described in details in \cref{a:low_C}, and leads to
\begin{align}
\langle \Phi(C)\rangle=1-C+C^2\left(2-\frac{16}{3\pi}\right)+O(C^3).
\end{align}
This expansion catches the behavior of $\langle \Phi(C)\rangle$
for small $C$, as seen in \cref{fig:low_C_behavior} of \cref{a:low_C}.

For larger values of $C$, the order parameter $\langle \Phi(C)\rangle$ decreases
and vanishes due to the transition. This decrease
becomes sharper and occurs at smaller values of $C$
when $\bar{A}_{syst}$ increases, as expected from
the decrease of $\Delta C_c$ and $C_c$ reported in \cref{fig:transition_average}.

In addition, the difference between the values of $\langle \Phi(C)\rangle$ 
for periodic and non-periodic boundary conditions
shown in  \cref{fig:phi_C_1real}(b)
is seen to decrease when the system size increases.
This suggests that the type of boundary condition
is asymptotically irrelevant for large systems,
and that our results for the transition threshold
obtained for large non-periodic systems should also provide an accurate
description of large periodic systems.

	\subsection{Clusters size distribution}
\label{s:size_distribution}

\begin{figure}[h!]
		\centering
		\includegraphics[width=0.87\linewidth]{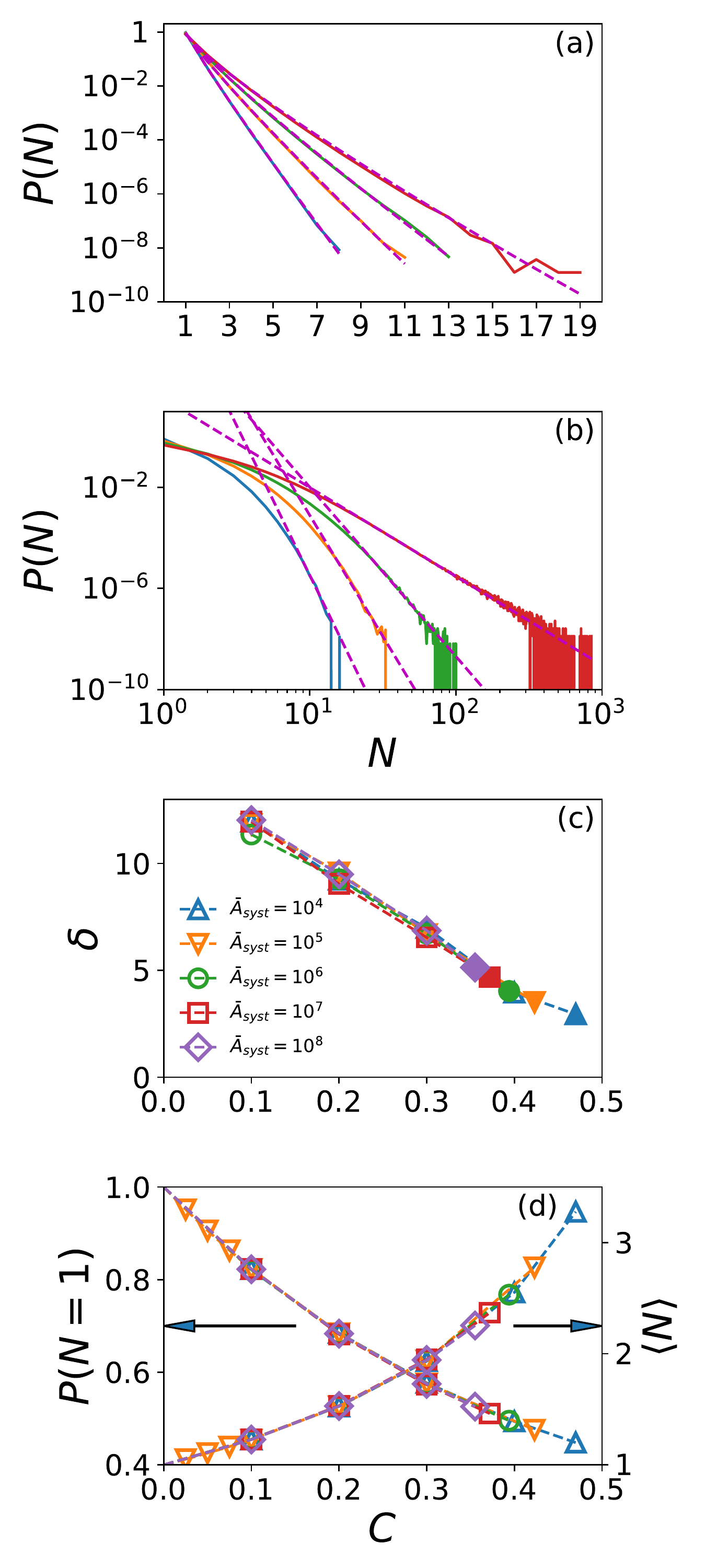}
		\caption{
  Cluster size distribution before transition.
        (a) $P(N)$ as a function of the number of discs $N$ in the cluster.
        Small values of $C$: from left to right $C=[0.025,0.05,0.075,0.1]$.
        System size: $\bar{A}_{syst}=10^5$, averaged over $10^5$ realizations. Dashed lines : fits with stretched exponentials
        $a_0\exp[-a_1(N-1)^{a_2}]$ with $a_2=[0.935,0.909,0.895,0.858]$.
        (b) $P(N)$ for larger values of $C$. Same system size, averaged over $10^4$ realizations 
		(the average is performed over realizations who did not experience a transition).
		From left to right $C=[0.1,0.2,0.3,0.423]$. 
		Dashed lines: power-law fits of the tails of distribution.
		(c) Exponent $\delta$ of the tail of the distribution $P(N)\sim N^{-\delta}$ as a function of $C$
		for various system sizes $\bar{A}_{syst}$.
		(d)	Probability of clusters with a single disc $P(N=1)$, and 
        average cluster size $\langle N\rangle$
		as a function of $C$ for various system sizes $\bar{A}_{syst}$. 
		Fixed boundary conditions.
        }
		\label{fig:cluster_size_distrib}
	\end{figure}

Let us call $N$ the number of discs in a cluster.
The probability distribution $P(N)$ before the transition,
is reported in \cref{fig:cluster_size_distrib}(a,b)~\footnote{
For each system size $\bar{A}_{syst} \in \{10^4,10^5,10^6,10^7,10^8\}$, we run $\{10^5,10^4,10^3,10^3,60\}$ realizations of our simulation. 
Then we extract the data for $C \in \{0.1,0.2,0.3,0.4,C_c=0.47\}$ for $\bar{A}_{syst}=10^4$ and $C \in \{0.1,0.2,0.3,C_c\}$ for $\bar{A}_{syst} \in \{10^5,10^6,10^7,10^8\}$, with $C_c \in \{0.423,0.394,0.372,0.355\}$. 
We exclude realizations where the system has already undergone a transition, 
corresponding to a significant number of runs mainly at $C=C_c$, where roughly half the realizations have transited. 
}.
When  $C\rightarrow 0$, there are mainly isolated discs
and  $P(N=1)\rightarrow 1$, as seen from \cref{fig:cluster_size_distrib}(d).
As shown in \cref{fig:cluster_size_distrib}(a), fit of $P(N)$ with a stretched exponential
($a_0\exp[-a_1(N-1)^{a_2}]$) shows that the distribution approaches an exponential distribution
when $C\rightarrow 0$, i.e. $a_2\rightarrow 1$ as $C\rightarrow 0$.

For larger values of $C$, the tails of the distributions for
large $N\gg1$ exhibit a power-law decay $P(N)\sim N^{-\delta}$.
The exponent $\delta$, reported in \cref{fig:cluster_size_distrib}(b), is  independent of $\bar{A}_{syst}$
and varies approximately linearly with $C$
\begin{align}
    \delta\approx 14.19-24.99 C\, .
    \label{eq:approx_delta}
\end{align}
Such a power-law tail of the distribution would lead
to a divergence of the average cluster size if $\delta\rightarrow 2$.
Using \cref{eq:approx_delta}, this condition suggests a divergence at $C\approx 0.49$. 
Note that this value is a speculative extrapolation in the sense that it is larger than all the values 
of $C$ reported in \cref{fig:cluster_size_distrib}.
The mean cluster size $\langle N\rangle$ is reported in \cref{fig:cluster_size_distrib}(c).
As expected, the mean cluster size $\langle N\rangle\rightarrow 1$ 
as $C\rightarrow 0$. However, the values of $C$ for which the 
transition can be measured do not allow one to probe the 
possible divergence of $\langle N\rangle$  at larger $C$.

	\subsection{Domain area distribution}
\label{s:area_distribution}

In systems such as those discussed in the introduction, 
the initial discs might not be observable, so that the 
cluster size $N$ might not be a relevant observable quantity.
Instead, the domain area could be observable.
As a reminder, a domain is the convex hull of 
all the discs contained in a cluster.
The normalized area of a domain is defined as
\begin{align}
    \bar A= \frac{A}{\pi r_d^2}
\end{align}
where $A$ is the domain area.

\begin{figure}[h!]
		\centering
		\includegraphics[width=0.9\linewidth]{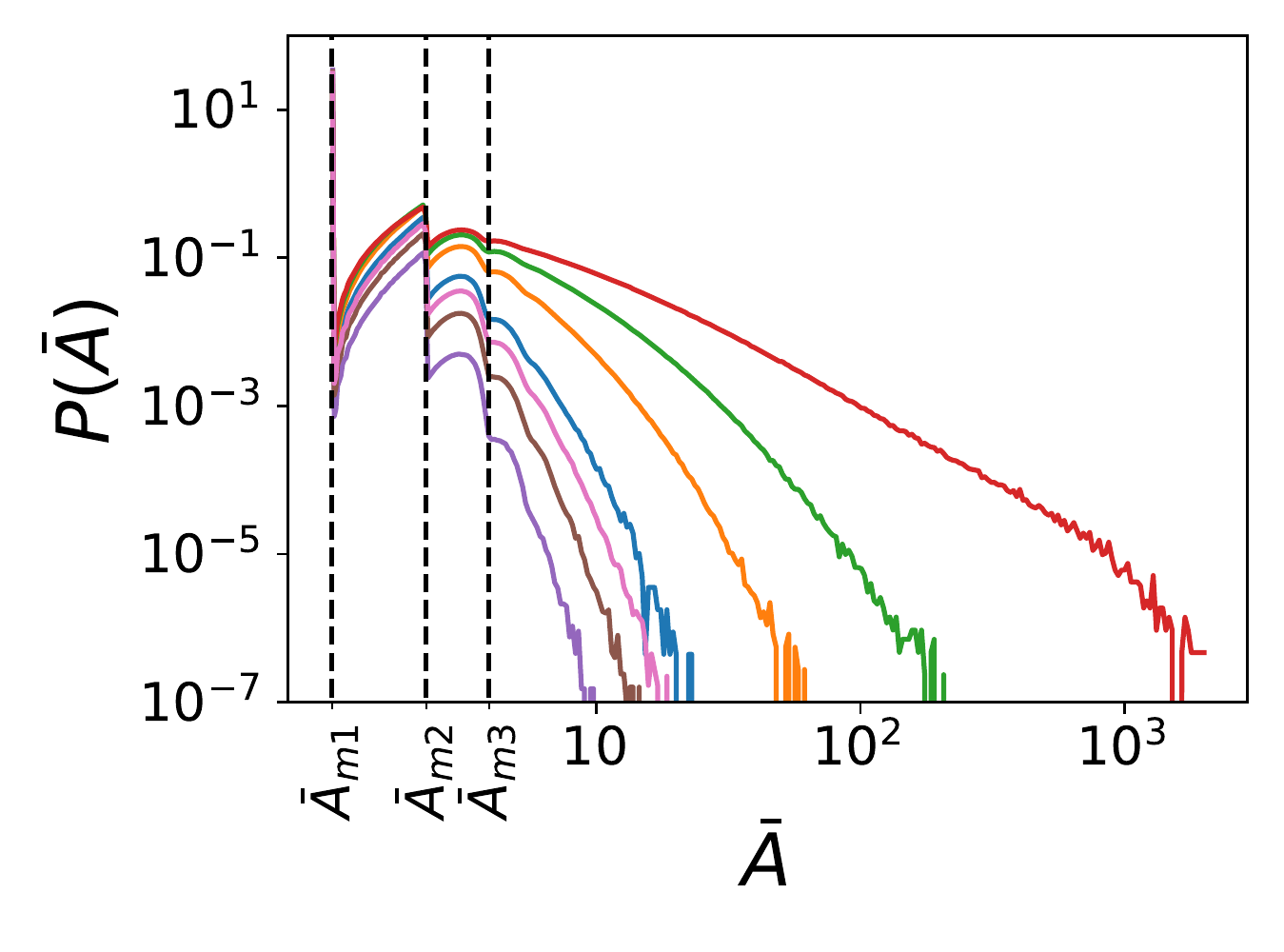}
		\caption{Clusters area distribution $P(\bar{A})$, for $\bar{A}_{syst}=10^5$, averaged
		over $10^4$ realizations, and $10^5$ realizations for the three curves on the left (small values of $C$), free boundary conditions. 
		From left to right: $C=[0.025,0.05,0.075,0.1,0.2,0.3,0.423]$
		(the last value of $C$ corresponds to the mean transition point $C_c$). 
    The special values $\bar{A}_{m1}$, $\bar{A}_{m2}$ and $\bar{A}_{m3}$ respectively correspond 
    to the maximal area of a cluster of one, two, and three discs, as discussed in the text.}
		\label{fig:PA}
	\end{figure}

The domain area probability distribution $P(\bar{A})$,
show in \cref{fig:PA}, is found to oscillate
at small $\bar{A}$. We observe several oscillations accompanied with 
discontinuities. First, a Dirac-delta-like peak
corresponding to the contribution of domains composed
of a single disc at $\bar{A}=\bar{A}_{m1}=1$. Then, clusters with two discs
only have a finite range of domain area $\bar{A}_{m1}<\bar{A}<\bar{A}_{m2}$,
and the distribution presents a step-like discontinuity
at $\bar{A}=\bar{A}_{m2}$. Possible areas of domains with 3 discs also obey
an inequality $\bar{A}_{m1}<\bar{A}<\bar{A}_{m3}$.
This leads to a slope-discontinuity in the area probability distribution
at $\bar{A}=\bar{A}_{m3}$ (in addition, we observe 
numerically that there is an approximately horizontal
tangent on the right side at $\bar{A}=\bar{A}_{m3}$).

The values of $\bar{A}_{mN}$ for small $N$ correspond
to the maximum domain area for a cluster of $N$ discs.
For two discs, the configuration with the largest area
is obtained for two tangent discs, leading to $\bar{A}_{m2}=1+4/\pi$.
For 3 discs, the configuration with the largest area
is obtained when two discs are tangent to the third disc,
and when the angle between the two tangent discs is $2\pi/3$,
leading to  $\bar{A}_{m3}=1+(4+3^{3/2})/\pi$.
As shown in \cref{fig:PA}, these values are in 
agreement with the observed singularities of $P(\bar{A})$.
The general question of the existence of higher order
$\bar{A}_{mN}$ and the nature of the associated series of singularities
remains open.

In contrast to $P(N)$, the tails of $P(\bar{A})$
for large $\bar{A}$ do not obey a clear power-law. This is rooted 
in the nonlinear relation between $A$ and $N$ which is discussed
in \cref{s:Appendix_numerical_methods}. Such a nonlinearity can be
interpreted as a finite-size effect for clusters
that are not large enough to obey the average-density
relation $N\approx \bar{A} C$.

	\subsection{Domain shape}
\label{s:cluster_shape}

\begin{figure}[h!]
	\centering
	\includegraphics[width=0.9\linewidth]{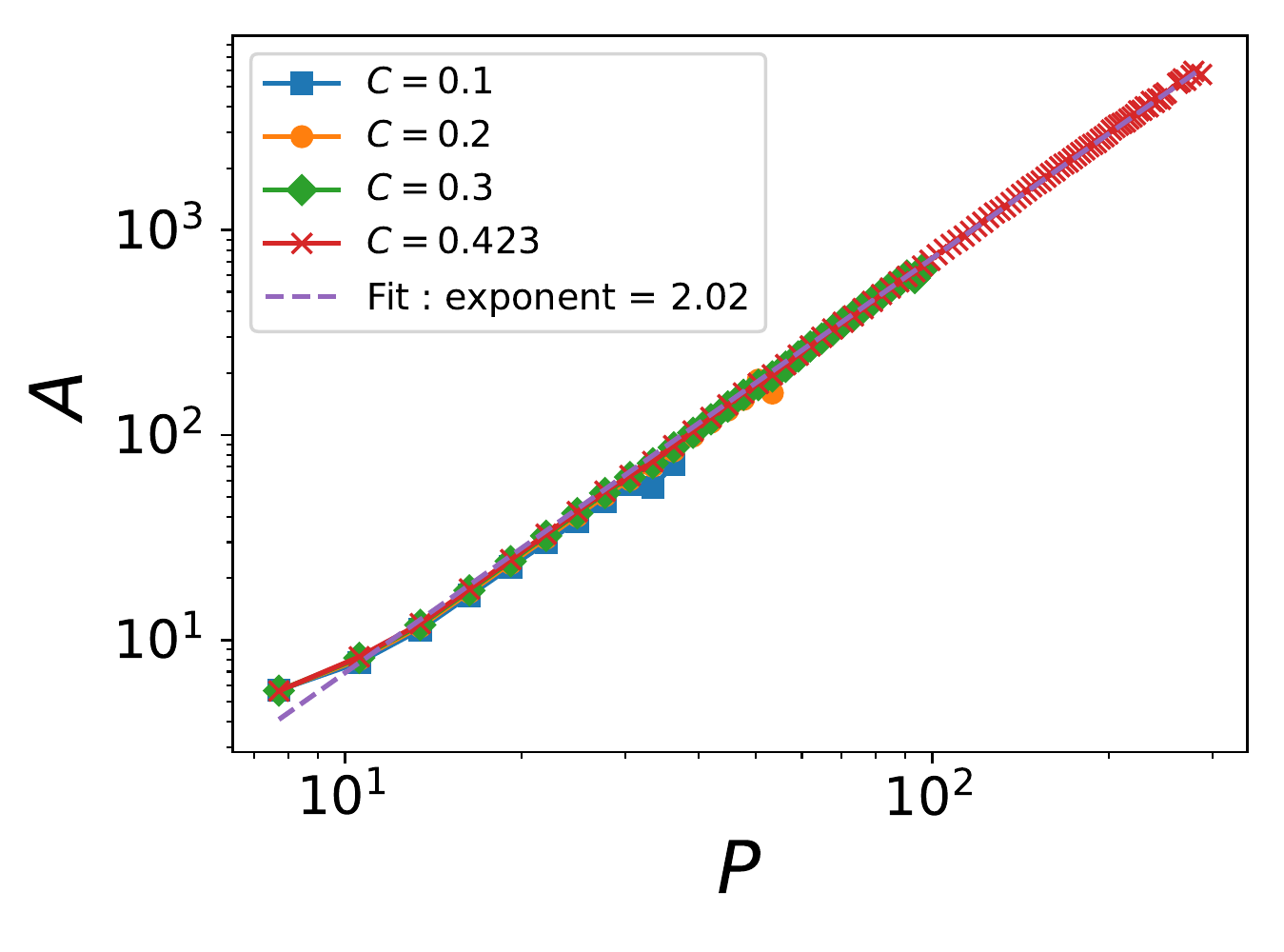}
	\caption{Average cluster area $A$ as a function of the perimeter $P$. 
	Same parameters as in \cref{fig:PA}.
    }
	\label{fig:A_P}
	\end{figure}

The shape of the domains can be
explored via their deviation from circularity.
In \cref{fig:A_P}, we see that the average domain area $A$ for a given perimeter $P$ 
is proportional to the square the domain perimeter $P^2$
for large domains.
For a circular domain, one has $4\pi A/P^2=1$.
For large domains, the prefactor is  constant and independent of $C$, and one finds $4\pi A/P^2\approx 0.827$.
This constant deviation from circularity for a wide range of sizes
suggests some scale-invariant properties of cluster shapes.
Such scale invariance can be associated to the presence of 
power-law tails of the cluster size distribution discussed in \cref{s:size_distribution}.

	\section{Discussion of the transition}
\label{s:discussion_transition}

\subsection{On-lattice bootstrap percolation} 
\label{s:comparison_boostrap_percolation}

\begin{figure}[h]
		\centering
		\includegraphics[width=0.9\linewidth]{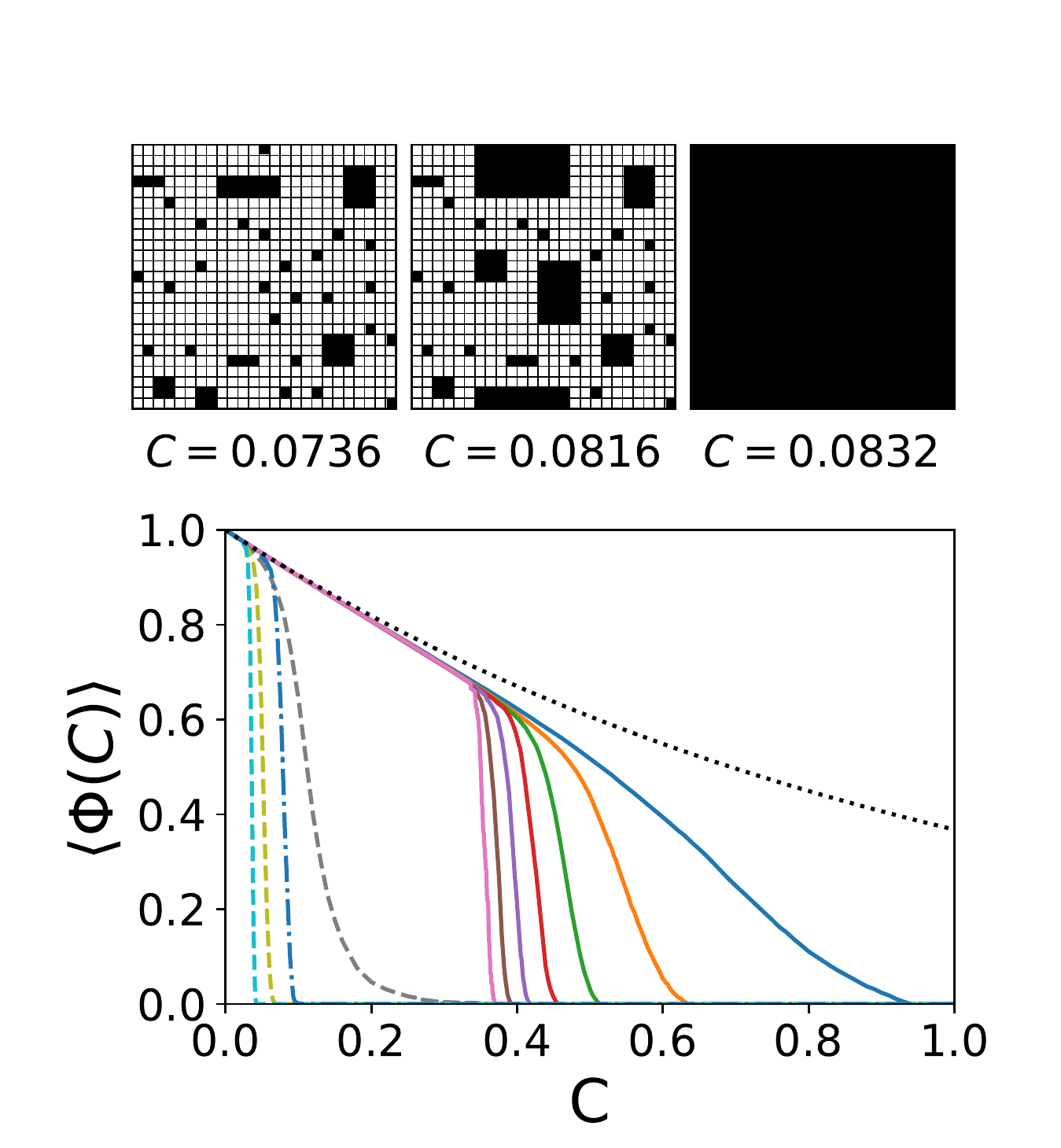}
		\caption{Comparison with bootstrap percolation.
		Upper panel: 
		final state of bootstrap percolation model on a square lattice 
        for three different values of 
		the initial concentration $C$. 
		Lower panel: order parameter $\langle\Phi(C)\rangle$ 
		our model (full lines), and for bootstrap percolation. Dashed lines : square lattice; from right to left : [$10\times10;10^2\times10^2;10^3\times10^3$] cells. Dash-dotted line : hexagonal lattice, $10^2\times10^2$ cells.}
		\label{fig:bootstrap_final}
		\label{fig:compare_boot_perc}
	\end{figure}

As discussed in the introduction, our model shares
strong similarities with bootstrap percolation models.
The bootstrap percolation model can be defined as follows.
Start with a lattice containing $N_{latt}$ cells.
Initially, $N_a$ cells are activated.
Then,  each inactive cell is made active if $m$ or more of its 
nearest neighbors are active. 
This activation process is performed iteratively
up to a final state where no new active cell can be added.
Final states corresponding to different values of $N_a$ are 
plotted in the top panels of  \cref{fig:bootstrap_final} for a periodic square
lattice with $m=2$ and $N_{latt}=25^2$. 
The results look similar to those of our model, and the system is invaded completely for large-enough $N_a$.

To compare bootstrap percolation with our results,
we need to define $C$ in bootstrap percolation.
Our procedure for the initial activation of the cells is designed to parallel
the continuum case. 
We choose a site randomly $N_i$ times.
Sites can be chosen more than one time.
If the chosen site is not active, we activate it. 
We then define
\begin{align}
    C=\frac{N_i}{N_{latt}}\, .
\end{align}
The probability of sites not to be activated initially
is $(1-1/N_{latt})^{N_i}\approx {\rm e}^{-N_i/N_{latt}}$
for $N_{latt}\gg 1$, and the average initial fraction of non-active sites is
\begin{align}
    \Phi_b(C)={\rm e}^{-C}\, ,
\end{align}
which is identical to \cref{eq:coverage_discs}.
The average number of initially active cells is then $N_a=N_{latt}[1-\Phi_b(C)]$.

A comparison of the numerical results of our model with those of bootstrap percolation
is reported in the lower panel of \cref{fig:compare_boot_perc}.
The behavior of the two models is similar as expected.
Both models give rise to a transition 
when $C$ is increased. However, the transition
occurs at much smaller values of $C$
for bootstrap percolation.

In order to gain further insight, we have also implemented
bootstrap percolation  on a hexagonal lattice~\cite{Adler1991}, 
with $m=3$ instead of $m=2$ for the square lattice
(indeed, one cell has $6$ nearest neighbors on a hexagonal lattice, instead of $4$ on a square lattice).
Due to their six-fold symmetry, hexagonal lattices can naively be considered
as more isotropic than square lattices.
As seen in \cref{fig:compare_boot_perc}, the bootstrap transition for a hexagonal
lattice is shifted to larger values of $C$ as compared to 
that of a square lattice, although the transition is still expected to
occur at $C\rightarrow 0$ in large systems~\cite{DeGregorio2021}. 
We speculate that this could be an indication that isotropy shifts
the transition to larger $C$ for finite system sizes. This trend would be consistent
with the observation of transitions at larger $C$ in our continuum model.

The bootstrap percolation transition on a square lattice
is known to obey~\cite{Zarcone1984,Adler1989,Holroyd2003}
\begin{align}
\label{eq:asymptotics_boostrap}
    \lim\limits_{\substack{%
	\Phi_b \to 1\\A_{syst}\to \infty}}(1-\Phi_b)\ln \bar{A}_{syst}=\frac{\pi^2}{9}\, .
\end{align}
Assuming that the convexification model can be viewed
as a continuous and isotropic limit of bootstrap percolation, 
a naive analogy based on the assumption that the microscopic
properties of the models are irrelevant at large scales would suggest that
$C_c\sim (\ln\bar{A}_{syst})^{-1} \rightarrow 0$ as $\bar{A}_{syst}\rightarrow\infty$ 
in the convexification model.

\subsection{Transition in the limit of small $C$ and large $\bar{A}_{syst}$} 
\label{s:transition_C_to_zero}

Our numerical results reported for the convexification model in \cref{s:transition} 
are obtained in systems that are too small 
to investigate a possible scaling 
regime similar to \cref{eq:asymptotics_boostrap}.
However, we can simply parallel the heuristic 
derivation of the asymptotic behavior as discussed in Refs.~\cite{Zarcone1984,Adler1989}.
The details of this analysis are reported in \cref{as:Analogy_Heuristic}.
The results are similar to those of usual on-lattice bootstrap
percolation and suggest that $C_c\sim (\ln \bar{A}_{syst})^{-2/3} \rightarrow 0$ as $\bar{A}_{syst}\rightarrow\infty$.

The only difference with on-lattice bootstrap percolation
is the exponent $2/3$. This difference comes from the fact that
the increase of cluster area $A$ due to merging with one active site
is proportional to the length of a facet of the cluster edge
 $\sim A^{1/2}$ in on-lattice bootstrap percolation,
while it is $\sim A^{1/4}$ for the merging of a cluster with
a single disc in the convexification model, as discussed in \cref{as:Analogy_Heuristic}. 
Further investigations would be
needed to clarify the scaling behavior as $\bar{A}_{syst}\rightarrow\infty$.

\subsection{Transition in at finite $C$} 
\label{s:transition_C_non_zero}

Experimental systems usually exhibit a finite size, and it is therefore
important to study the behavior of the transition for finite $C$.
Inspired by the heuristic derivation of Refs.~\cite{Zarcone1984,Adler1989},
we propose again a heuristic analysis to probe the transition.
One difficulty in the analysis of the finite concentration
regime comes from the relevance of clusters of all sizes
while we could assume that the system was mainly composed
of isolated discs in the limit $C\rightarrow 0$.

Following Refs.~\cite{Zarcone1984,Adler1989},
our analysis is based on the assumption that the 
largest cluster is the one that invades the system.
The statistics of the largest cluster is studied
in \cref{a:extreme_finite_C}.
We account for the presence of power-law tails
of $P(N)$ reported in \cref{s:size_distribution}, 
and the predictions are in good agreement with the simulations.

The analysis reported in details in \cref{a:transition_heuristic}
is based on a scenario in two steps. First, the largest cluster
starts to grow, and then this growth produces a macroscopic
avalanche which spans the whole system.
The results suggests that, as in on-lattice bootstrap percolation, 
the transition is governed by the propagation of a macroscopic avalanche.

This analysis reproduces qualitatively the decrease
of the transition threshold with $C$, but predicts a
threshold which is quantitatively lower.
Moreover, no true scaling regime emerges in this regime. 
We attribute this absence of scaling
to the absence of linear relation between the number $N$
of discs in a domain and the domain area $\bar{A}_N$ in clusters of finite size. 
However, is is possible to fit the decrease of the threshold using
a power-law, and we find that $C_c\sim \bar{A}_{syst}^{-\alpha_c}$
with $\alpha_c\approx 0.032$. 
Using a fit of the simulation results in the range  
$10^6\leq \bar{A}_{syst}\leq 10^8$,
we obtain a similar to but smaller exponent $\alpha_c\approx 0.024$.

To summarize this section, our simple model inspired from Refs.~\cite{Zarcone1984,Adler1989} 
captures the qualitative behavior of the transition, but 
is unable to reach quantitative agreement.

	\section{Comparison with experimental results on graphene de-adhesion}
\label{s:experiments}

In this section, we wish to compare our results with the
experimental data from \cite{Yamamoto2012} on the de-adhesion
of graphene caused by the intercalation of nanoparticles.
Following \cite{Yamamoto2012}, the radius of the detachment
zone around a single isolated nanoparticle with diameter
$d$ can be approximated 
by the radius of detachment associated to a vertical point force
lifting the membrane up to a distance $d$ from the substrate.
Solving the Föppl-von K\'arm\'an equations\cite{Yamamoto2012,Schwerin1929,Komaragiri2005}, one obtains
	\begin{equation}	
		2R=\left(\frac{4nG}{3\gamma}\right)^{\frac{1}{4}} d ,
		\label{eq:radius_experim}
	\end{equation}
where 
$nG$ is the tensile rigidity of the $n$-layer graphene, 
and $\gamma$ is the adhesion energy per unit area of graphene on silica.
Using the experimental values reported in Ref.~\cite{Yamamoto2012}	
$d=7.4\pm 2.2\ nm$, $G=2.12\times 10^3\ eV/nm^2$,
and	$\gamma=1.7\pm 1.1\ eV/nm^2$, we find
$R\approx 23.6n^{1/4}$nm. 
Moreover, we have $C=\rho\pi R^2$, where $\rho$ is the density of nanoparticles. 
In the experiments of Ref.\cite{Yamamoto2012}, $\rho=160\pm 24\ \mu m^{-2}$.
Inserting these values into \cref{eq:radius_experim}, we obtain $C= \alpha n^{\sfrac{1}{2}}$, where $\alpha=0.28\pm0.16$.
Note the large uncertainty on $\alpha$,
which is dominated by the uncertainty on the particle diameter.
A fit of the experimental data for the experimental order parameter $\Phi_{exp}$ at small $C$ prior to transition 
where $\Phi(C)\approx 1-C\approx1-\alpha n^{1/2}$
leads to $\alpha\approx0.18$,
which is consistent with the estimate of $\alpha$ reported above. 
	
\begin{figure}[h!]
	\centering
	\includegraphics[width=0.9\linewidth]{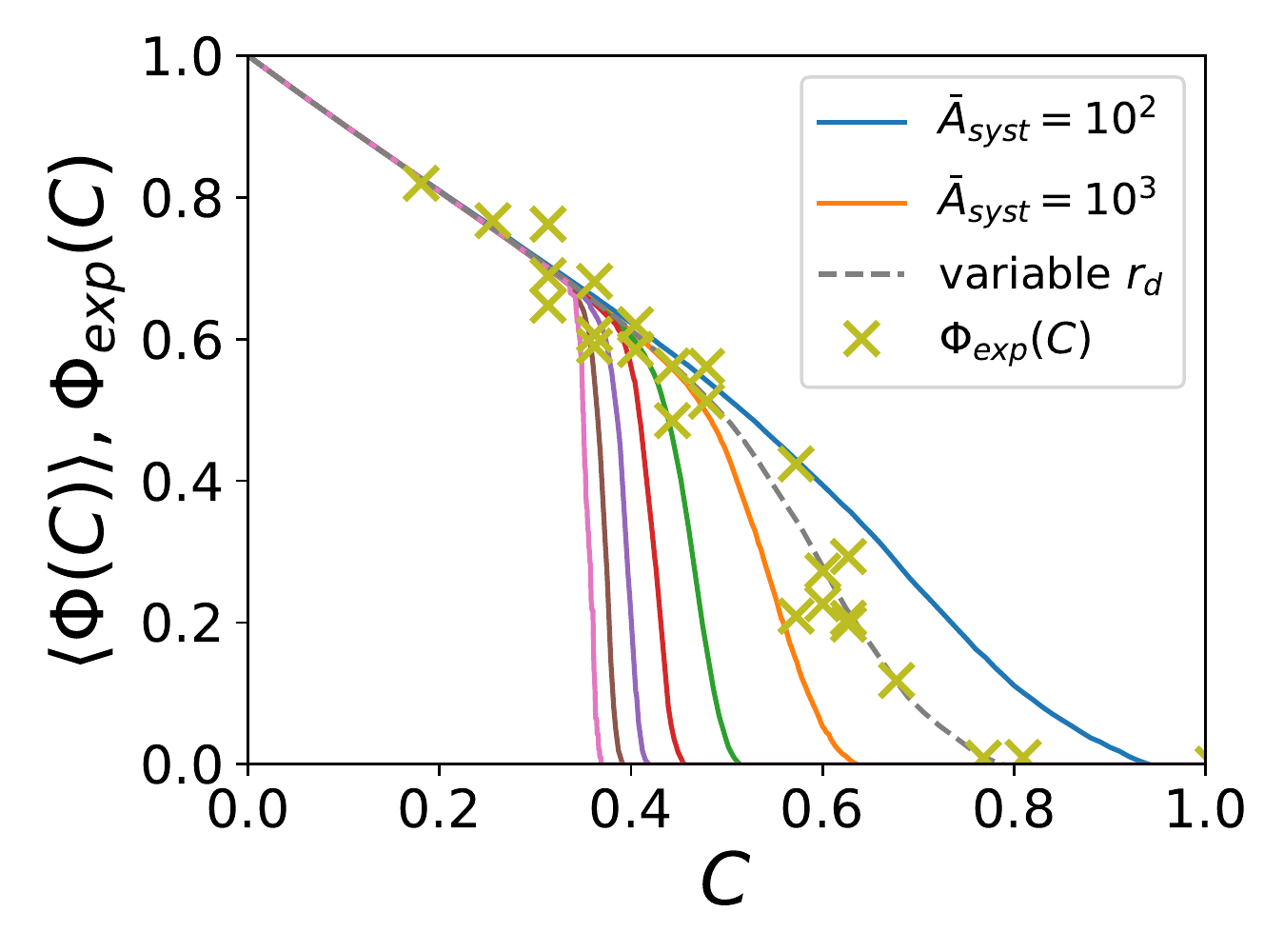}
	\caption{Comparison of the convexification model with experimental de-adhesion data.
	Full lines : simulations where $C$ is varied by adding discs one by one.
	Dashed lines: simulations with variable $r_d$.
	Crosses : experimental data of \cite{Yamamoto2012}. 
    }
	\label{fig:comparison_experiments_intercal}
\end{figure}

An inspection of \cref{fig:comparison_experiments_intercal} suggests
that the transition observed for experimental data corresponds 
to a system size $\bar{A}_{syst}$ between $10^2$ and $10^3$.
However, in experiments, the number $n$ of layers of graphene varies, 
and therefore from \cref{eq:radius_experim}, 
the radius $r_d$ of individual detachment discs varies
while the density $\rho$ of particles remains constant.
To model these conditions, 
we have performed simulations where $r_d$ varies while
the number of discs $N_d$ and the system area $A_{syst}$ remain constant.
The disc radius $r_d$ is varied from $0$ to $r_{dmax}=[A_{syst}/(\pi N_d)]^{1/2}$ leading to 
a variation of $C=r_d^2/r_{dmax}^2$ from $0$ to $1$.
Here, the system size $A_{syst}$ is only setting the 
scale of all lengths via $r_{dmax}$.
The result of the simulations therefore only depend on the two dimensionless
numbers $N_d$ and $C$, as opposed to the previous
simulations with dimensionless numbers $\bar{A}_{syst}$ and $C$.
As shown in \cref{fig:comparison_experiments_intercal},  agreement with experiments 
is found for $N_d\approx 2\times 10^2$.
Additional discussions about the variable $r_d$ 
transition curves are reported in \cref{a:variable_rd}.

Our results therefore suggest a finite size effect 
involving about $N_d=200$ particles in experiments.
Translating $N_d$ into a lengthscale defined as the diameter
of a circular zone including 200 discs, we find $2(N_d/\pi\rho)^{1/2}\approx 1.3\mu$m.
However, the physical mechanism at the origin of this lengthscale is not known.
For example, correlations between nanoparticle positions,
or elastic effects could come into play.

	\section{Conclusion}
\label{s:conclusion}
In conclusion, we have introduced a model for
invasion in two-dimensions based on the convexification
and merging of domains. Starting with an initial
condition where discs of equal diameter are randomly placed in the plane, 
we find an invasion transition. 
An analogy with on-lattice bootstrap percolation suggests that
the invasion threshold should be observed at zero density when the system size
tends to infinity. 
Our numerical simulations do not allow us 
to investigate the asymptotic behavior for very small disc densities.
We hope that our work will motivate 
further studies of the convexification model. 
Indeed, theoretical investigations and large scale  numerical simulations
are needed to elucidate the low-density and large scale
asymptotic properties of the model.

In the regime of finite densities which should be the relevant one 
for most experimental systems, 
we found that the cluster size distribution
has a power-law tail, with an
exponent that increases linearly with the disc density.
In this regime, the deviation from circularity of the average shape of large clusters is constant. 

Furthermore, in both finite and small concentration regimes, 
we found that the domain area distribution oscillates for small areas.
These oscillation are accompanied with singularities of the 
distribution.

Our results compare favorably with experimental data
on the unbinding transition of graphene with 
intercalated nanoparticles reported in Ref.~\cite{Yamamoto2012}. Other experimental
applications of the model would be welcome to strengthen the 
claim of genericity and universality of domain invasion via convexification.

\acknowledgements
We wish to thank Mahito Yamamoto for providing
the experimental data from \cite{Yamamoto2012} used in \cref{fig:comparison_experiments_intercal},
Dajiang Liu for enlightening discussions,
Tristan Pierre-Louis for remarks on \cref{a:periodic_BC}
and Fran\c{c}ois Detcheverry for a critical reading of
the manuscript.

	\begin{appendix}

	\section{Details on the simulation algorithm}
\label{a:numerics}

Below, we provide some details about the 
algorithm that we have used to build the clusters.

\begin{figure}[h!]
    \centering
    \includegraphics[width=0.65\linewidth]{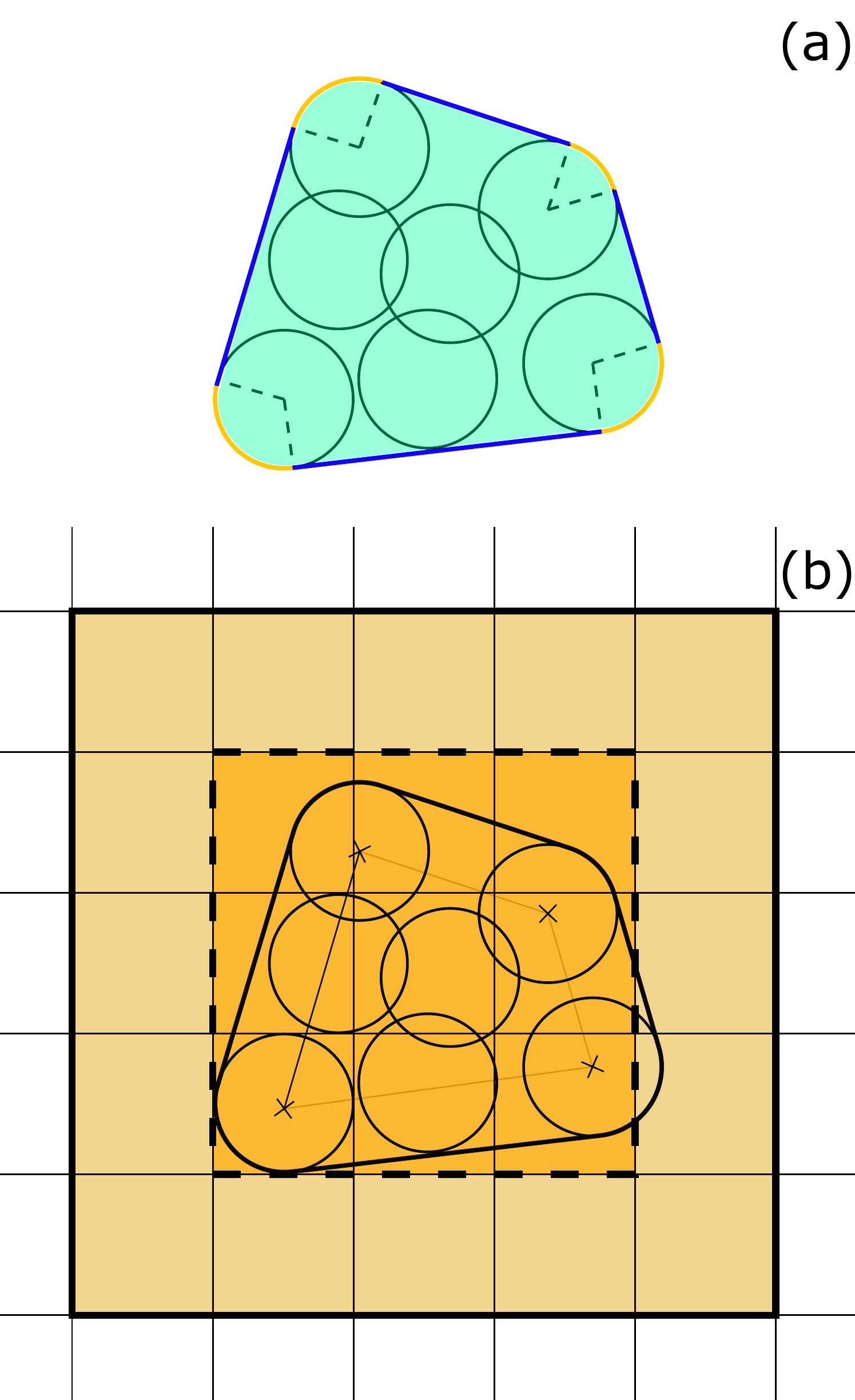}
    \caption{Schematic of a seven-disc clusters with four outlying discs. 
    (a)
    The light (green) shading represent the cluster domain.
    The boundary of the domain is composed of straight segments and
    arcs of circle.
    (b)
    Compartimentalization grid with squares of size $2r_d\times 2r_d$.
    A rectangle is constructed, which contains all discs centers plus
    one shell of grid cells (shown in lighter shading).
    Contact tests are performed with other discs only when their center fall
    into this rectangle. 
    }
    \label{fig:covering_mesh}
\end{figure}

For each cluster, we use the Jarvis march
algorithm to obtain the lists containing the $N_{out}$ outlying discs that are
in contact with the boundary of the domain.
As shown in \cref{fig:covering_mesh}, the boundary of the domain is composed of
arcs of circles that belong to the edges of the outlying discs,
and segments between two outlying discs.
Discs inside the cluster (i.e., not outlying) can be discarded
for the computation of the merging and convexification processes.

We add a discs one by one.
A new disc will belong to a given cluster if the center of the new disc
falls inside the fictitious convex-hull domain of the cluster that is formed by assuming discs of radius $2r_d$.
Testing the contact requires the computation of $N_{out}$ tests for checking
on which side of the domain boundary segments the center of the new disc is,
and $N_{out}$ tests for checking if the center of the new disc is in the outlying discs.

If the new disc is is contact with no cluster, a new one-disc cluster is created.
Otherwise,
the new disc is merged with the first cluster it is found to be in contact with. 

If the new disc is a new outlying disc of the cluster it is merged with, then
the new disc is added to the list of outlying discs. In such a case the cluster
domain changes and grows, and
the algorithm starts testing the contact of this modified cluster with the other clusters. 
To test the contact between the growing cluster and the other clusters,
we test the contact of each outlying disc of other clusters with the growing cluster. 
From this point on, at each step of convexification, 
the growing cluster is simultaneously merged with all the clusters that are in contact with it. 
Then, the new convex hull is computed, and contact tests are resumed.
The growth process is iterated until no contact is found between the clusters.

To accelerate the simulations, a
compartmentalization method is implemented, following Ref.~\cite{Gawlinski1981}. 
The system is covered by a square grid with cells of side $2r_d$. 
Contact tests of a disc
with a cluster are performed only if the center of the disc
belongs to the minimal rectangle of grid cells that contains the centers of the discs of the cluster
plus one shell of cells, as shown in \cref{fig:covering_mesh}.

As the growing cluster becomes large, 
the compartmentalization method tends to be counter-productive 
and slows the simulations  down.
This happens mostly during the macroscopic avalanche at the transition. 
Hence, if the number of outlying discs of the growing cluster exceeds
an empirically chosen value of 20, compartmentalization
is not used.

	\section{Periodic boundary conditions}
\label{a:periodic_BC}

	\begin{figure}[h!]
		\centering
		\includegraphics[width=\linewidth]{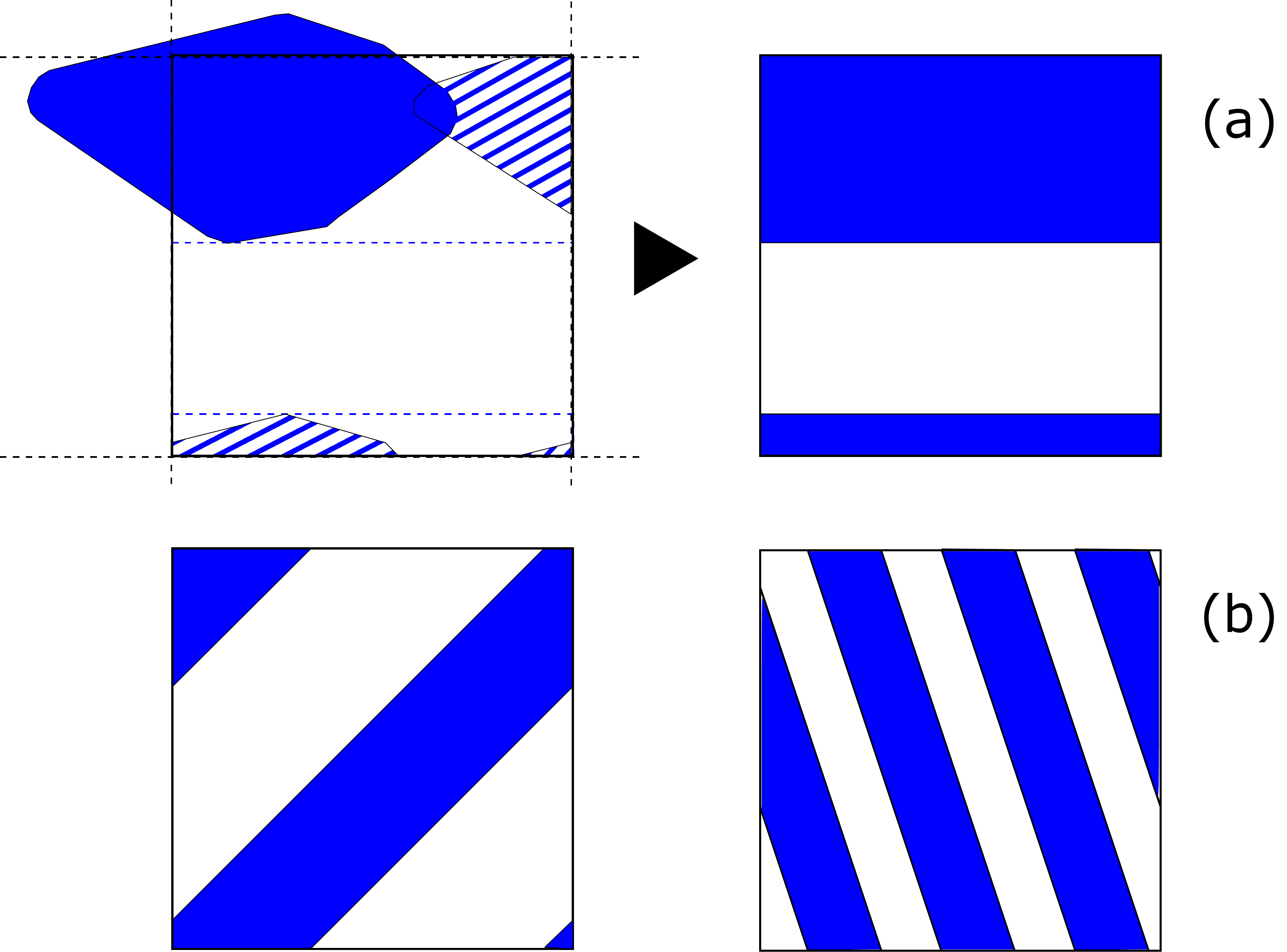}
		\caption{
		(a) The convex hull of a cluster
		overlapping with its image through the periodic boundary
		conditions is a strip-like domain.  (b) The strip may wrap around the system
		an arbitrary number of times in both directions.
        }
		\label{fig:overlap_PBC}
	\end{figure}

	In the periodic case, the algorithm is similar to that described above.
	However, a novel situation appears when a domain is in contact with its periodic image,
	as shown in \cref{fig:overlap_PBC}(a). 
	The domain is then made convex only along the direction where it is in contact with its periodic image,
	leading to a strip-like convex-hull.
	To include all possible cases, since our square periodic simulation box has a torus-like topology,  
	we have to consider situations where the domain
	wraps around the boundary conditions several times,
	as shown on \cref{fig:overlap_PBC}(b).
	
	Strips are convex domains in the sense that, given two points belonging to the strip, 
	it is possible to find a segment that connects these two points, with all points of the segment also belonging to the strip. 
	However in contrast to usual convex shapes in the plane, for a given strip, 
    it is possible to find a segment connecting two points of the strip
    with a part of the segment not belonging to the strip.
    More precisely, strips appear when defining 
    a convex domain as a domain that contains all the straight segments
    between two points of the domain
    if these two points can be merged to one point by a continuous
    transformation that keeps them inside the domain.
    Equivalently,
    strips can be defined by moving to the full
    infinite plane, tiled by our square system 
    and all its periodic images and taking the convex hull 
    of each connected components.

    Periodic systems have the advantage of not being
    influenced by boundary conditions as in the case of fixed boundaries. 
    However, the periodicity itself imposes discrete
    global symmetries that make the strips anisotropic. These symmetries here 
    are those of the square tiling of the plane, and the related anisotropy is 
    expressed in the fact that the orientation of the strips cannot take
    arbitrary values and are limited to rational slopes in the $x,y$ plane.
      
 	In addition, when the domain is in contact with its periodic image in two
	orthogonal directions, the convex hull is the whole simulation box. 	
 	In practice for large system size, each time the addition of a disc provoked 
    the appearance of a strip, 
 	the process of iterative convexification continued 
    and led to the full invasion of the domain by this strip.

\section{Low-$C$ expansion for $\langle \Phi(C)\rangle$ }
\label{a:low_C}

In the following, we report on two expansions of $\langle \Phi(C)\rangle$
in the limit  $C\rightarrow 0$.

\subsection{Expansion from  area occupied by the sum of all disc areas}

The first strategy is to use
the situation where each disc contributes to decreasing 
the free area by their own area $\pi r_d^2$ as a leading order estimate. 
This leads to a first estimate of the fraction of the system
area that is not covered by the domains 
\begin{align}
    \langle \Phi(C)\rangle_{a0} =1-C.
    \label{eq:all_0}
\end{align}
In the limit of low density $C\rightarrow0$, 
this expression is valid to linear order in $C$ because
the density of discs is very low so that the probability
that discs overlap vanishes. 

A first correction to this approximation is to consider
the effect of dimers, i.e. clusters made of two overlapping discs.
Overlapping occurs if another disc is present at a distance
lower than $2r_d$. For each dimer, the area not covered by the 
domains must be corrected by an amount
\begin{align}
    \Delta A= \pi r_d^2-2r r_d 
\end{align}
where $r$ is the distance between the centers
of the two discs. Hence, 
in the presence of a concentration $\rho_d$ of discs, the correction is
\begin{align}
    \langle \Phi(C)\rangle_{a1}=\frac{1}{2} \rho_d \int_0^{2r_d}dr 2\pi r \rho_d\Delta A
    =C^2 \left(2-\frac{16}{3\pi}\right).
        \label{eq:all_1}
\end{align}
where the $1/2$ prefactor accounts for the fact that each dimer
is counted two times (one time starting from the first disc, and another time 
starting from the second disc).
The correction $\langle \Phi(C)\rangle_{a1}$ is positive, to wit:
on average over all distances $r$, the positive contribution coming from
the overlapping area of the discs is larger than the negative contribution
due to the convexification.

\subsection{Expansion from the empty area}

Another strategy is to perform an 
expansion from the area that is empty 
in the presence of the overlapping discs
before any convexification is performed.
We then start with 
\begin{align}
    \langle \Phi(C)\rangle_{e0} =\Phi_p(C)={\rm e}^{-C},
        \label{eq:empty_0}
\end{align}
which is consistent with \cref{eq:all_0} to linear order in $C$.

A dimer correction is then obtained 
considering a different area change that has to account for the fact
that overlapping of discs is already present in the reference state
\begin{align}
    \Delta A=  r_d^2\left(
    \pi-2u+u(1-\frac{u^2}{4})^{1/2}-2\arccos\frac{u}{2}
    \right)
\end{align}
where $u=r/r_d$.
We then obtain
\begin{align}
    \langle \Phi(C)\rangle_{e1}=\frac{1}{2}\rho_d \int_0^{2r_d}dr 2\pi r \rho_d\Delta A
    =C^2 \left(\frac{3}{2}-\frac{16}{3\pi}\right).
    \label{eq:empty_1}
\end{align}
In contrast to \cref{eq:all_1}, the correction $\langle \Phi(C)\rangle_{e1}$ is now negative because it 
only results from the negative contribution of convexification.

The 
Taylor expansion of the two approximations is identical to up second 
order in $C$, i.e.
\begin{align}
    \langle \Phi(C)\rangle&=
    \langle \Phi(C)\rangle_{a0}+\langle \Phi(C)\rangle_{a1}+O(C^3)
    \nonumber \\
    &=\langle \Phi(C)\rangle_{e0}+\langle \Phi(C)\rangle_{e1}+O(C^3)
    \nonumber \\
    &=1-C+C^2 \left(2-\frac{16}{3\pi}\right)+O(C^3).
    \label{aeq:expansion_low_C}
\end{align}

\subsection{Comparison of the low-$C$ expansions with numerical results}

\begin{figure}
    \centering
    \includegraphics[width=\linewidth]{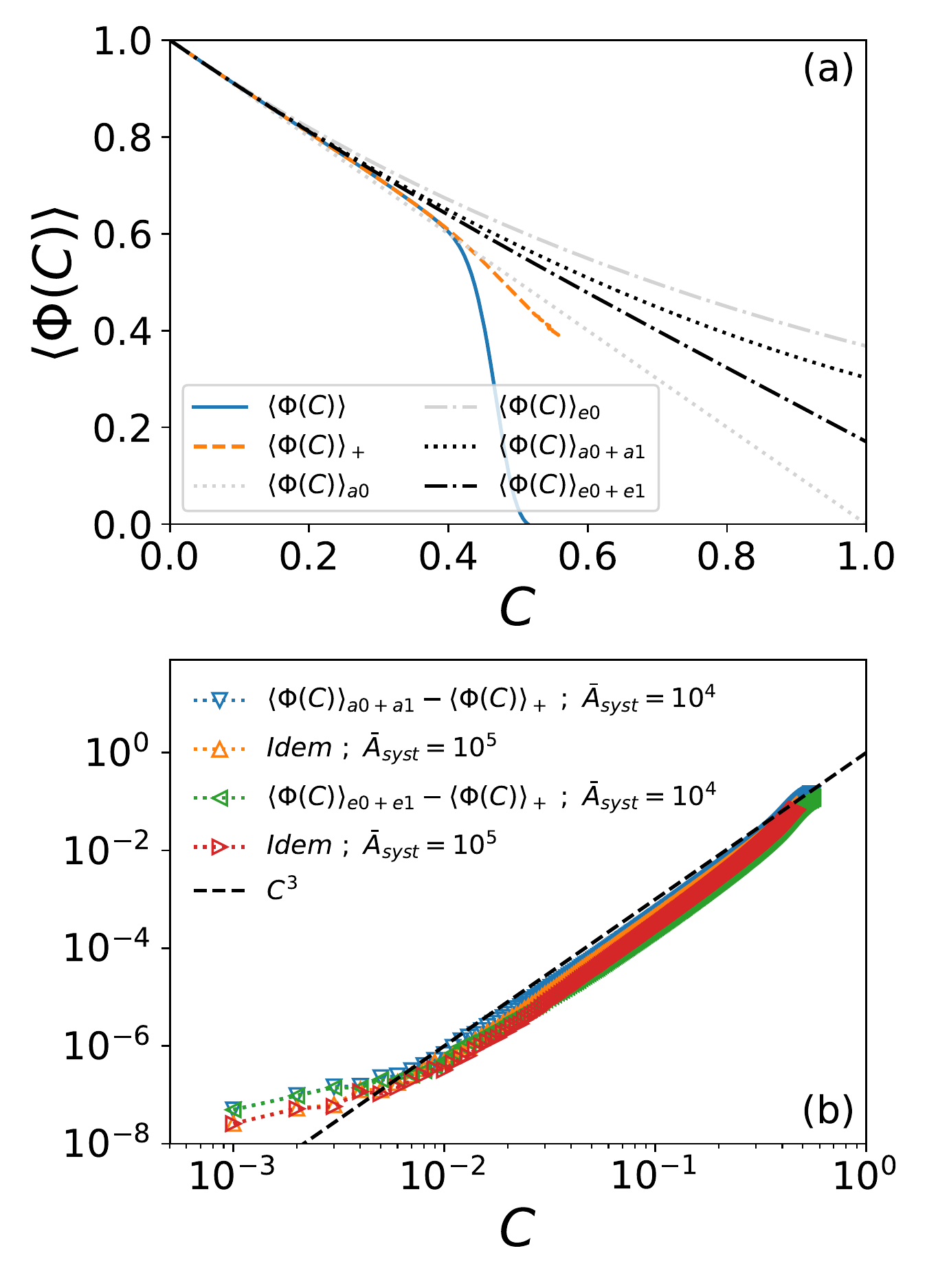}
    \caption{
    Low $C$ expansion of the order parameter $\langle \Phi(C)\rangle$.
    (a) Comparison with simulations for $\bar{A}_{syst}=10^4$.
    Solid line $\langle \Phi(C)\rangle$ from simulations. 
    Dashed line, average of $\Phi(C)$ in systems that have not undergone
    a transition $\langle \Phi(C)\rangle_{+}$. Dotted lines:
    first expansion $\langle \Phi(C)\rangle_{a0}$ and 
    $\langle \Phi(C)\rangle_{a0+a1}=\langle \Phi(C)\rangle_{a0}+\langle \Phi(C)\rangle_{a1}$.
    Dashed-dotted lines: second expansion $\langle \Phi(C)\rangle_{e0}$ and 
    $\langle \Phi(C)\rangle_{e0+e1}=\langle \Phi(C)\rangle_{e0}+\langle \Phi(C)\rangle_{e1}$.
    The leading order expressions are shown in lighter tone (gray).
    (b) Deviation of simulation results from low-$C$ predictions.
    }
    \label{fig:low_C_behavior}
\end{figure}

As seen in \cref{fig:low_C_behavior}(a), low $C$ expansion of $\langle \Phi(C)\rangle$ are in good agreement
with numerical simulations at small $C$. Numerical results 
for the average $\langle \Phi(C)\rangle_+$ in systems that have not 
undergone a transition is also reported. This latter quantity
deviates from $\langle \Phi(C)\rangle$ in the transition region.

The deviation from the expansions, reported in \cref{fig:low_C_behavior}(b),
scales as $C^3$ as expected from \cref{aeq:expansion_low_C}.
Some deviations are observed for $C$ smaller than $10^{-2}$.
These deviations are caused by the finite size of the system
that does not present enough dimers at small $C$ for the expansions to be valid.

\section{Variable $r_d$}
\label{a:variable_rd}

In the main text, 
the transition is analyzed by adding discs one by one, i.e. by varying $N_d$.
In terms of the dimensionless parameters $(C,\bar{A}_{syst})$, this corresponds to varying $C$ 
at fixed $\bar{A}_{syst}$.
Another equivalent procedure is to fix $N_d$ and $A_{syst}$,
and to increase the disc radius $r_d$.
This procedure is less convenient for numerical simulations,
but could be relevant for comparison with experimental system
like those reported in \cref{s:experiments}.

\begin{figure}[h!]
	 	\centering
	 	\includegraphics[width=\linewidth]{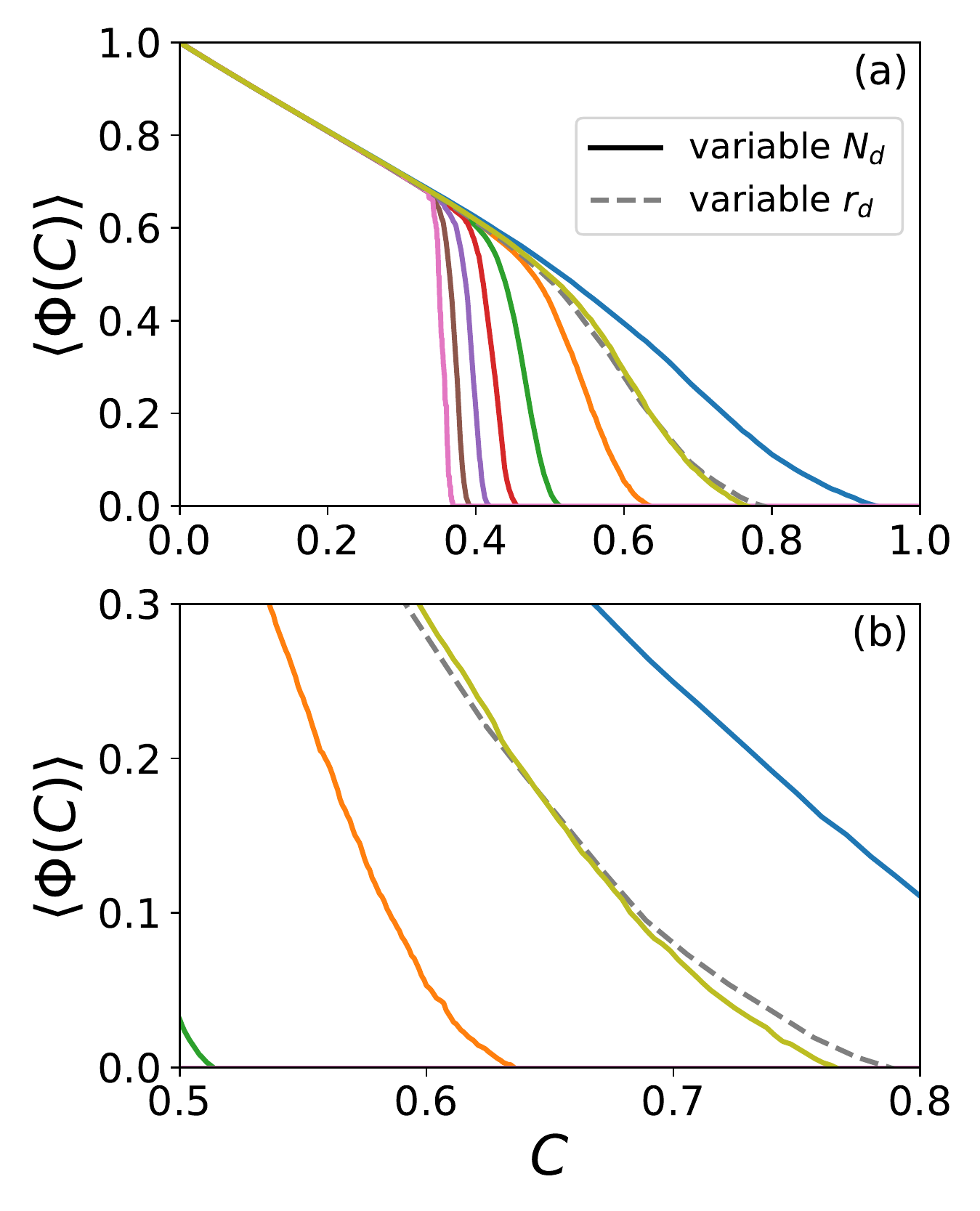}
	 	\caption{$\langle \Phi(C)\rangle$ for variable $r_d$. 
   Same parameters as in \cref{fig:phi_C_1real}.
    Full lines: we explore $C$ by adding discs one by one, i.e., by varying $N_d$ for a constant radius of discs $r_d = 1$. 
    Dashed line: we explore $C$ by varying $r_d$, for a fixed number of discs $N_d=200$ (we keep ${A}_{syst}=200\pi$ fixed and $r_d$ varies from $0$ to $1$, so that $\bar{A}_{syst}$ varies from $+\infty$ to $200$). 
   To see that the variable $r_d$ curves cross the variable $N_d$ curves, we also show a variable $N_d$ curve for $\bar{A}_{syst} = 307.7$ (light-green curve). }
	 	\label{fig:range_rd}
	 \end{figure}

Increasing $r_d$ corresponds to increasing $C$ 
and decreasing $\bar{A}_{syst}$ simultaneously.
Hence, increasing $r_d$ can be seen as continuously 
switching to curves with lower and lower values of $A_{syst}$ when increasing $C$.
However, we observe in \cref{fig:range_rd} that the curve $\langle \Phi(C)\rangle$ 
as a function of $C$
for a given value of $\bar{A}_{syst}$ is always below 
the $\langle \Phi(C)\rangle$ curve for a lower value of $\bar{A}_{syst}$.
Hence, the variable $r_d$ procedure leads to smaller
slopes for the decrease of $\langle \Phi(C)\rangle$ as a function of $C$.
Indeed, we see that the variable $r_d$ curves have a smaller slope and cross the variable
$N_d$ curves in \cref{fig:range_rd}.
As a consequence, the width a the transition along $C$ is slightly larger
with the variable $r_d$ procedure.

\section{Analogy with heuristic discussion of 
on-lattice bootstrap percolation at small $C$}
\label{as:Analogy_Heuristic}

In this section, we apply the heuristic derivation
of the asymptotic behavior of the transition for on-lattice bootstrap
percolation~\cite{Zarcone1984,Adler1989} to the case of 
our convexification model.

Consider a large domain. In the limit $C\rightarrow 0$,
there are very few large clusters, and we therefore analyze the growth
of a cluster by merging with isolated discs.
The probability not to grow $P^-({ A})$ for a domain
of perimeter ${ A}$ is the probability
for having no disc in a region 
of width $r_d$ around the periphery of the domain.
Denoting the area of this region as ${ A}_e$,  we have
\begin{align}
    P^-({ A})={\rm e}^{-{ A}_e\rho_d}.
\end{align}
In normalized variables  $\bar{A}_e={ A}_e/(\pi r_d^2)$
\begin{align}
    P^-(\bar{A})={\rm e}^{-\bar{A}_eC}.
\end{align}
Hence, in the limit $C\rightarrow 0$, the probability to grow is
\begin{align}
    P^+(\bar{A})=1-{\rm e}^{-\bar{A}_eC}\approx \bar{A}_eC.
    \label{eq:P^+_lin}
\end{align}
Assuming that the cluster is large and circular of radius $R$,
we have
\begin{align}
    \bar{A}_e=\frac{{ A}_e}{\pi r_d^2}
    =\frac{2\pi R r_d}{\pi r_d^2}
    =2\frac{(A/\pi)^{1/2}}{r_d}
    =2\bar{A}^{1/2}.
\end{align}
Growth of the largest cluster of area $\bar{A}_{c}$
at the critical concentration $C_c$ requires $P^+(\bar{A}_c)=a$
where $a$ is a constant, leading to the condition
\begin{align}
    2\bar{A}_{c}^{1/2}C_c\approx a.
    \label{eq:condition_a_critical}
\end{align}

Let us now investigate the growth of the cluster via
mergings with discs. 
Consider first the merging of a cluster with one disc.
Assuming that clusters are circular with radius $R$, the increase
of the cluster area due to merging is 
\begin{align}
\Delta A=(R^2-r_d^2)(\tan\theta-\theta)
\end{align}
where $\cos\theta=(R-r_d)/d$, with $d$ the distance between the center
of the cluster of radius $R$ and the disc.
In the limit of interest  where $R\gg r_d$, we have 
\begin{align}
\Delta A=\frac{2^{3/2}}{3} R^{1/2}\left(1+\alpha\right)^{3/2} r_d^{3/2}
\label{eq:Delta_A}
\end{align}
where $\alpha=(d-R)/r_d$ obeys $-1\leq \alpha\leq 1$.
Assuming a uniform distribution of $\alpha$ in the interval $[-1,1]$,
we approximate the term $(1+\alpha)^{3/2}$ by its the average value 
\begin{align}
\left(1+\alpha\right)^{3/2}\approx \frac{1}{2}\int_{-1}^{1}d\alpha \left(1+\alpha\right)^{3/2}= \frac{2^{5/2}}{5}.
\label{eq:average_alpha}
\end{align}

The normalized area of the cluster after the $n$-th mergings obeys
\begin{align}
    \bar{A}_n=\bar{A}_{n+1}+\Delta \bar{A}_n
\end{align}
From \cref{eq:Delta_A,eq:average_alpha}, and 
assuming that all clusters are circular, we have
\begin{align}
    \Delta \bar{A}_n= \frac{2^4}{15\pi}\bar{A}_n^{1/4}.
\end{align}
Taking the continuum limit in $n$, and integrating over $n$, we obtain
\begin{align}
    \bar{A}_n= \left\{\frac{2^2}{5\pi}(n+n_0)\right\}^{4/3}.
    \label{eq:Delta_Abar_n}
\end{align}
where $n_0$ is a constant that will be neglected in the limit
of large $n$.

The probability to have a cluster of size $\bar{A}_n$
obeys the recursion relation
\begin{align}
    P( \bar{A}_{n+1})=P^+(\bar{A}_{n}) P( \bar{A}_{n}).
\end{align}
Combining this relation with \cref{eq:P^+_lin} leads to
\begin{align}
    P( \bar{A}_{n+1})=2\bar{A}_{n}^{1/2}C P( \bar{A}_{n}).
\end{align}
Hence, using \cref{eq:Delta_Abar_n}, we have
\begin{align}
    P( \bar{A}_{n})=P(1) \prod_{m=1}^{n-1} 2C\left\{\frac{2^2}{5\pi}m\right\}^{2/3}.
\end{align}
Assuming that $P(1)\approx C$, we obtain
\begin{align}
    P( \bar{A}_{n})=C^n ((n-1)!)^{2/3}\left\{\frac{2^{7/3}}{(5\pi)^{2/3}}\right\}^{n-1}.
\end{align}
We then impose the condition that there is a 
cluster of size $\bar{A}_{n_c}$ with a finite probability $b$ at the threshold
\begin{align}
    \bar{A}_{syst}P( \bar{A}_{c})=b.
    \label{eq:condition_b_critical}
\end{align}
Using the Stirling formula and \cref{eq:Delta_Abar_n}, this condition leads to
\begin{align}
    C_c (2\pi(n_c-1))^{1/3}
    \left\{\frac{2}{\mathrm {e}^{2/3}} C_c\bar{A}_c^{1/2}(1-\frac{1}{n_c})^{2/3}\right\}^{n_c-1}
    \!\!\!=\frac{b}{\bar{A}_{syst}}.
\end{align}
Using \cref{eq:condition_a_critical}, and assuming again $n_c\gg 1$, we obtain
\begin{align}
    C_c (2\pi n_c)^{1/3}
    \left\{\frac{a}{\mathrm {e}^{2/3}} \right\}^{n_c}
    \!\!\!=\frac{b}{\bar{A}_{syst}}.
\end{align}
Hence to leading order
\begin{align}
    n_c \approx \frac{\ln\bar{A}_{syst}}{\ln [\mathrm {e}^{2/3}/a]}.
\end{align}
Using again \cref{eq:Delta_Abar_n,eq:condition_a_critical} this 
relation is rewritten as
\begin{align}
    C_c \approx \frac{a}{2}\left(
    \frac{2^2}{5\pi}\frac{\ln [\mathrm {e}^{2/3}/a]}{\ln\bar{A}_{syst}}\right)^{2/3}.
\end{align}
Hence, we conclude that 
$C_c\sim (\ln\bar{A}_{syst})^{-2/3}\rightarrow 0$ 
as $\bar{A}_{syst}\rightarrow 0$.

\section{Link between cluster number and disc density}
\label{a:relations_notations}

In this appendix, we derive some relations on the average number of clusters,
and we define some quantities that will be used in the following appendices.

The total number of discs $N_d$ obeys
\begin{align}
\sum_{N=1}^\infty N M_N=N_d,
\end{align}
where $M_N$ is the average number of clusters of size $N$.
Since $P(N)=M_N/M_{cl}$, where $M_{cl}$ is the total number of clusters, we have
\begin{align}
M_{cl}=\frac{N_d}{\sum_{N=1}^\infty N P(N)}.
\label{eq:Mtot}
\end{align}
Defining the moments
\begin{align}
\mu_k=\sum_{N=1}^\infty N^k P(N),
\label{eq:mu_n}
\end{align}
we rewrite \cref{eq:Mtot} as
\begin{align}
M_{cl}=\frac{N_d}{\mu_1}=\frac{C\bar{A}_{syst}}{\mu_1}.
\label{eq:Mtot_mu1}
\end{align}

\section{Largest cluster size in the finite $C$ regime}
\label{a:extreme_finite_C}

In this appendix, we propose an expression for the largest cluster size $N_{max}$.
The analysis is based on the numerical observation of a power-law behavior
for the cluster size distribution $P(N)\approx B N^{-\delta}$ for large $N$.

The distribution $P(N)$ is approximated with an ansatz of the form
\begin{align}
P(N)=\frac{B}{(N_0+N)^{\delta}}.
\end{align}
where $N_0>-1$ is a constant.
The normalization of $P(N)$
\begin{align}
\sum_{N=1}^{\infty}P(N)=1,
\label{eq:norm_P(N)}
\end{align}
imposes 
a condition that allows one to determine $N_0$
\begin{align}
B\,\zeta[\delta,N_0+1]=1
\label{eq:N0_dist}
\end{align}
where $\zeta$ is the Hurwitz Zeta function
~\footnote{The Hurwitz Zeta function is defined as
$$
\zeta(s,a)=\sum_{n=0}^\infty \frac{1}{(n+a)^s}.
$$
}.

A simple approximation for the average largest cluster size $N_{max}$
when choosing $M_{cl}$ clusters from the distribution $P(N)$
is obtained by the condition that only one cluster has a size
larger than $N_{max}$~\cite{Krapivsky2010}. Assuming that $N_{max}\gg 1$, we 
can use the continuum limit
\begin{align}
\int_{N_{max}}^{\infty}dN\,P(N)=\frac{1}{M_{cl}}.
\label{eq:cond_extreme_value}
\end{align}
Using \cref{eq:Mtot_mu1}, we obtain
\begin{align}
N_{max}=\left(\frac{C\bar{A}_{syst}B}{(\delta-1)\mu_1}\right)^{1/(\delta-1)}
-N_0,
\end{align}
where
\begin{align}
\mu_1=B\,\zeta[\delta-1,N_0+1]-N_0.
\label{eq:mu1_N0}
\end{align}

\section{Heuristic discussion of the transition in the finite $C$ regime}
\label{a:transition_heuristic}

In the following, we present a heuristic derivation
for the expression for the transition probability.

Upon the addition of a new disc, 
the probability that a macroscopic avalanche
occurs is assumed to be the product of two terms:
the probability $P_{start}$ that the growth of the cluster
of largest size $N_{max}$ can be triggered,
and $P_{av}$ the probability that this initial growing cluster
gives rise to a macroscopic avalanche that spans the whole system.
The probability $Q_{trans}(N_d)$ that no transition has occurred 
before $N_d$ therefore obeys 
\begin{align}
Q_{trans}(N_d+1)&=Q_{trans}(N_d)(1- P_{start} P_{av}),
\label{aeq:Qtrans_discrete}
\end{align}

In the following, we obtain expressions for $P_{start}$ and $P_{av}$,
neglecting corrections due to the system boundary.

\subsection{Probability trigger an avalanche from the merging of a disc with a cluster}

In the following, we derive an expression for
$P_{start}$ based on the assumption that growing
domain of size $\sim N_{max}$ is triggered by
the deposition of a disc on the largest domain.
The probability $P_{start}$ is therefore the product
of the probability $P_{dep}$ that a newly deposited disc 
leads to a modification of the edge of the largest cluster,
and of the probability that
subsequent growth of the cluster edge occurs 
after the adding of the first disc.

The probability  $P_{dep}$ is the probability
that a new disc is deposited in 
a zone of width $2r_d$ along the edge of the largest domain,
leading to an initial growth of the domain
\begin{align}
P_{dep}=\frac{4\pi R_{max}r_d}{A_{syst}}=\frac{4 \bar{R}_{max}}{\bar{A}_{syst}}.
\label{eq:pdep}
\end{align}
The radius $R_{max}$ of the largest cluster is defined from the relation
\begin{align}
\pi R_{max}^2=A_{N_{max}},\;\mbox{or}\;\;\; \bar{R}_{max}^2=\bar{A}_{N_{max}},
\end{align}
where $N_{max}$ is given by  \cref{eq:cond_extreme_value}.

In the following, we derive an approximate expression for the 
probability that the initial growth of the cluster 
due to the deposition of a new disc gives rise to 
subsequent growth by merging with other clusters.

\begin{figure}[h!]
	 	\centering
        \includegraphics[width=0.8\linewidth]{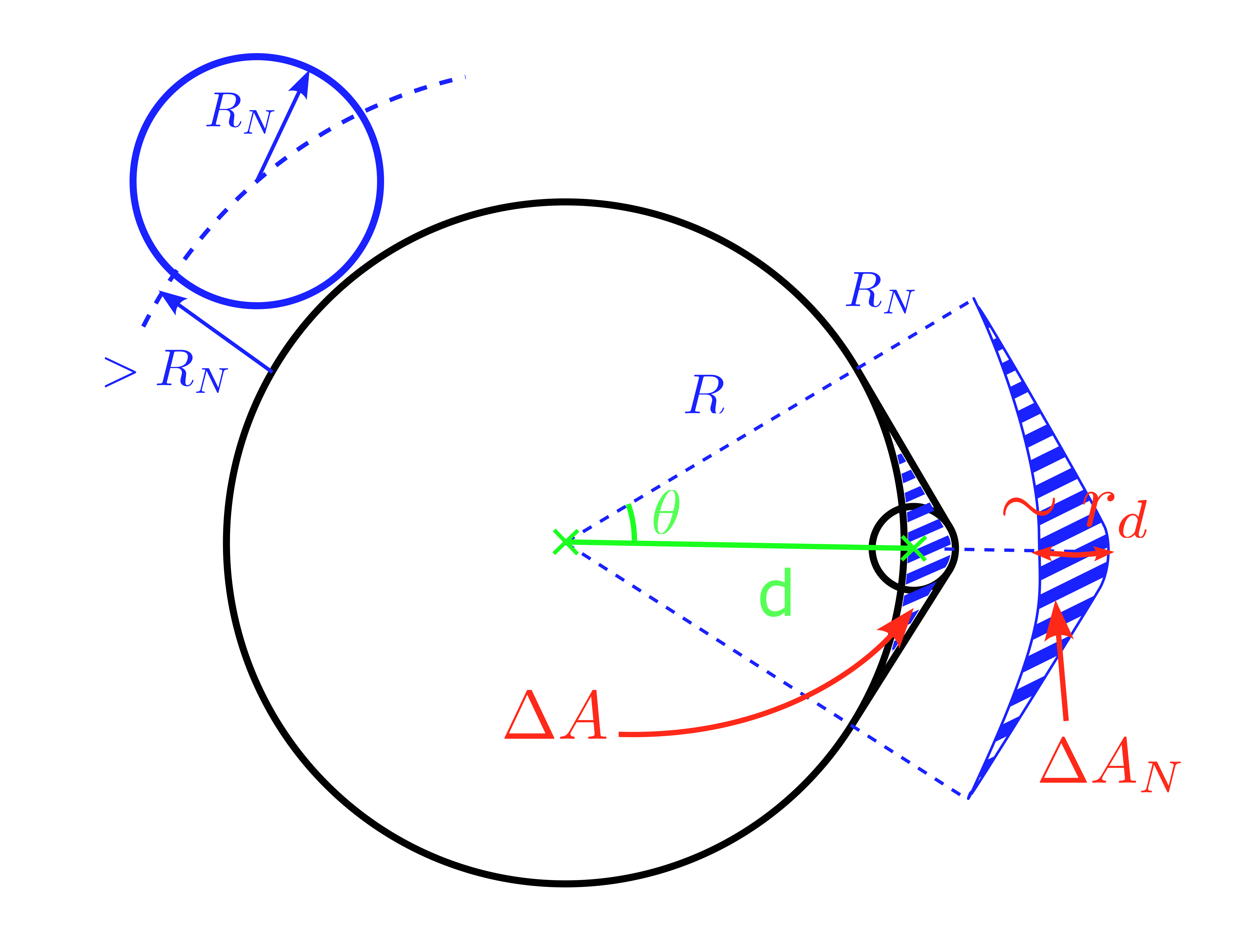}
	 	\caption{Schematic for the derivation of the probability
   that the growth of the cluster due to the deposition of 
   a new disc triggers further merging with other clusters.
   }
    	\label{fig:schematic_strat}
	 \end{figure}

We first simplify the analysis by assuming that all clusters are circular.
The probability that a domain of radius $R$ does not grow further
after the initial growth due to a newly added disc at its edge is
\begin{align}
Q_+(R)=\exp\left[-\sum_{N=1}^\infty \rho_N \Delta A_N\right]
\label{aeq:Q+_sum_DeltaAN}
\end{align}
where $\rho_N$ is the number density of clusters of size $N$,
and $\Delta A_N$ is the area of the region where the center
of a merging cluster of radius $R_N$ can be placed. 
This zone, which is constrained by the fact that the domains
of size $R_N$ initially do not overlap with the domains of radius $R$, is represented in \cref{fig:schematic_strat}.
While the width of this region in the direction perpendicular to the initial position
of the edge does not depend on the radius $R_N$ of the merging domain,
the width in the other direction is actually proportional to $R+R_N$.
Hence, we have 
\begin{align}
\Delta A_N\approx \frac{R+R_N}{R} \Delta A,
\label{aeq:DeltaAN_DeltaA}
\end{align}
where $\Delta A$ is the area of the initial growth region due to 
the merging of the domain of radius $R$ with the deposited disc.
Inserting \cref{aeq:DeltaAN_DeltaA} in \cref{aeq:Q+_sum_DeltaAN}
and using \cref{eq:Delta_A,eq:average_alpha}, we obtain
\begin{align}
Q_+(R)=\exp\left[-\frac{2^4}{15}\, R^{1/2} r_d^{3/2}
\sum_{N=1}^\infty \rho_N \left(1+\frac{R_N}{R}\right).
\right]
\end{align}
Using the relation \cref{eq:Mtot_mu1}, we have
\begin{align}
\pi r_d^2\rho_N=\pi r_d^2\frac{M_N/M_{cl}}{A_{syst}/M_{cl}}= \frac{C P(N)}{\mu_1}.
\end{align}

We now define 
\begin{align}
\tilde{N}=A_N\rho_d=\bar{A}_NC=\pi R_N^2\rho_d=\bar{R}_N^2C.
\label{eq:disc_density_circular_domain}
\end{align}
where $\bar{R}=R/r_d$.
Since the disc density inside clusters is expected to be constant
and equal to $\rho_d$ for large clusters, we have
$\tilde{N}\rightarrow N$ as $N\rightarrow\infty$.

Finally, we obtain 
\begin{align}
Q_+(\bar{R})=\exp\left[-\frac{2^4}{15\pi\mu_1}\left(
\bar{R}^{1/2} C+ \frac{C^{1/2}}{\bar{R}^{1/2}}\tilde{\mu}_{1/2}
\right)\right]
\label{eq:proba_cascading}
\end{align}
with
\begin{align}
\tilde{\mu}_k=\sum_{N=1}^\infty \tilde{N}^k P(N).
\label{eq:mu_nt}
\end{align}

Finally, the probability  
that the domain of largest size starts to grow, 
which is approximated by the probability that the 
first growth gives rise at least to one merging event with another cluster reads
\begin{align}
P_{start}=P_{dep}(1-Q_+(\bar{R}_{max})).
\end{align}

\subsection{Probability to propagate a macroscopic avalanche}

We now derive an expression for $P_{av}$, 
the probability that the growth of the domain 
proceeds indefinitely after the starting regime.
The difference between the starting regime and further growth is
essentially due to the constraint of non-penetration of the clusters.
In the initial stages of the avalanche, this forbids the presence
of clusters of size $N$ at a distance smaller than their radius $R_N$
from the cluster. In the later stages of growth, the 
edge of the growing domain is not stationary and can be in contact with other
clusters~\footnote{ Note however that we neglect the correlations betwen the clusters
other than the growing cluster in all cases.}.

To evaluate $P_{av}$, we start with the probability
of a large growing circular non-stationary domain of radius $R$ not to grow,
which is the probability that no other domain of size $N$ is in a strip of width
$2R_N$ around the edge of the domain. For $R_N\ll R$, the area of the strip is $4\pi R R_N$,
and we have
\begin{align}
Q_{stop}(R)=\exp\left[
-\sum_{N=1}^\infty \rho_N 4\pi R R_N 
\right].
\end{align}
The probability to grow indefinitely is the product of the probability $1-Q_{stop}(R)$
for all $R$s that are reached during the growth process.
Hence
\begin{align}
\ln P_{av} &=\sum_R \ln [1-Q_{stop}(R)]
\nonumber \\
&\approx \int_{R_{max}}^\infty \frac{dR}{\Delta R} \ln [1-Q_{stop}(R)] ,
\end{align}
where $\Delta R$ is the change of radius due to a merging event.
 Since we will consider the largest cluster,
the other clusters are smaller and one can safely use \cref{eq:Delta_A} with the 
substitution $r_d\rightarrow R_N$ for the change of area induced by a single merging event. Hence, the 
change in radius due to one merging is
\begin{align}
\Delta R=\frac{\Delta A}{2\pi R}=\frac{2^4}{15\pi }\frac{R_N^{3/2}}{R^{1/2}}.
\end{align}
Finally, we obtain
\begin{align}
P_{av} &=\exp\left[ \frac{15\pi}{2^7}
\frac{\mu_1^{3/2}}{\tilde\mu_{3/4}\tilde\mu_{1/2}^{3/2}} 
\int_{x_{max}}^\infty \hspace{-0.3 cm} dx\, x^{1/2}\ln [1-{\rm e}^{-x}] 
\right],
\end{align}
with 
\begin{align}
x_{max}&=\frac{C^{1/2}\tilde\mu_{1/2}}{\mu_1}\bar{R}_{max}.
\end{align}

\subsection{Transition probability}

Changing variable from $N_d$ to $C$ in \cref{aeq:Qtrans_discrete}, 
 taking the continuum limit, and using the relation $\mathrm d N_d/\mathrm d C=\bar A_{syst}$, 
we obtain a differential equation
\begin{align}
\frac{\mathrm d}{\mathrm dC}Q_{trans}(C)=-4 \bar{R}_{max}
(1-Q_+(\bar{R}_{max}))P_{av}Q_{trans}(C),
\end{align}
which is solved with the initial condition $Q_{trans}(0)=1$ as
\begin{align}
Q_{trans}(C)=\exp\left[-4
\int_0^C\hspace{-0.3 cm}\mathrm d C \, \bar{R}_{max} (1-Q_+(\bar{R}_{max}))P_{av}\right].
\end{align}
The numerical evaluation of $Q_{trans}(C)$ 
for some given system size $\bar{A}_{syst}$ requires to 
evaluate the quantities  $\mu_1$, $\tilde{\mu}_{1/2}$
and $\tilde{\mu}_{3/4}$ that depends on $C$,
and $\bar{R}_{max}$ that depends on $C$ and $\bar{A}_{syst}$.

\subsection{Numerical methods}
\label{s:Appendix_numerical_methods}

For the numerical evaluation of $Q_{trans}(C)$, we use a strategy based
on parametric analytical formulas 
that fit the numerical results for $\delta(C)$, $B(C)$ and $\bar{A}_N$ in the range
of $0.1\leq C \leq C_c$ where we have measured 
$P(N)$ and $A_N$ numerically. 
These analytical formulas  have no theoretical basis.
However, we report in the main text the strikingly simple
linear behavior of the exponent $\delta(C)$.
As an important remark, the precise form of these formulas should not
influence our results since they are essentially used for interpolation.
In contrast, the power-law behavior $P(N)\sim N^{-\delta}$ is crucial
since it has been used for extrapolation to 
large $N$ in \cref{eq:cond_extreme_value} to determine
the size of the largest cluster $N_{max}$.

\begin{figure}[h!]
	 	\centering
	 	\includegraphics[width=\linewidth]{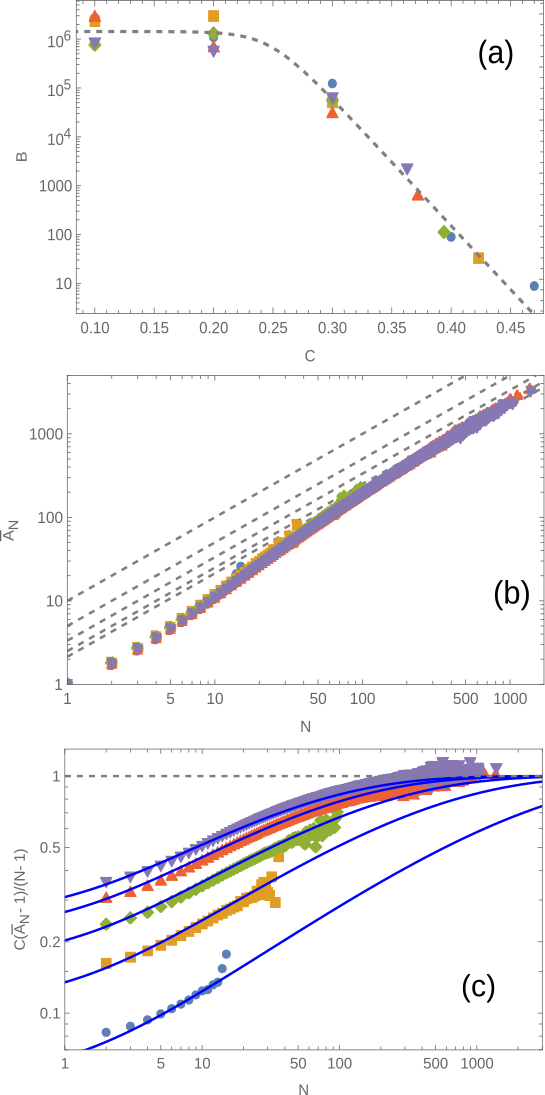}
	 	\caption{(a) Parameter $B$ extracted from simulation data as a function of $C$.
   Circles, squares, diamonds, up-triangles and down-triangles
   correspond to system sizes $10^8,10^7,10^6,10^5$ and $10^4$ respectively.
   (b) Cluster area $A_N$ as a function of the number 
   of discs $N$ for $\bar{A}_{syst}=10^5$.
   Circles, squares, diamonds, up-triangles and down-triangles
   correspond to $C=0.1$, $0.2$, $0.3$, $0.4$ and $0.468$.
   Dashed lines are $\bar{A}_N=N/C$.
   For concentrations lower than $0.4$, the asymptotic behavior $A_N\approx N/C$
   is not reached in our simulations.
   (c) Normalized area distribution $C(\bar{A}_N-1)/(N-1)$.
   Lines are the comparison to \cref{eq:ansatz_AN}.
   }	 	\label{fig:results_analytical}
	 \end{figure}

The first step in these evaluations is to determine
the dependence of $\delta$ and  $B$ on $C$.
The variation of $\delta$ with $C$ is given by \cref{eq:approx_delta}.
In addition, we use 
\begin{align}
B= \frac{b_2}{1+{\mathrm e}^{b_0 (C-b_1)}}.
\label{eq:ansatz_B}
\end{align}
As seen \cref{fig:results_analytical}(a) this expression catches the 
variation of $B$ with $C$ when 
$b_0=60$, $b_1=0.25$ 
and $b_2=1.4\times 10^6$.

\begin{figure}[h!]
	 	\centering
	 	\includegraphics[width=0.9\linewidth]{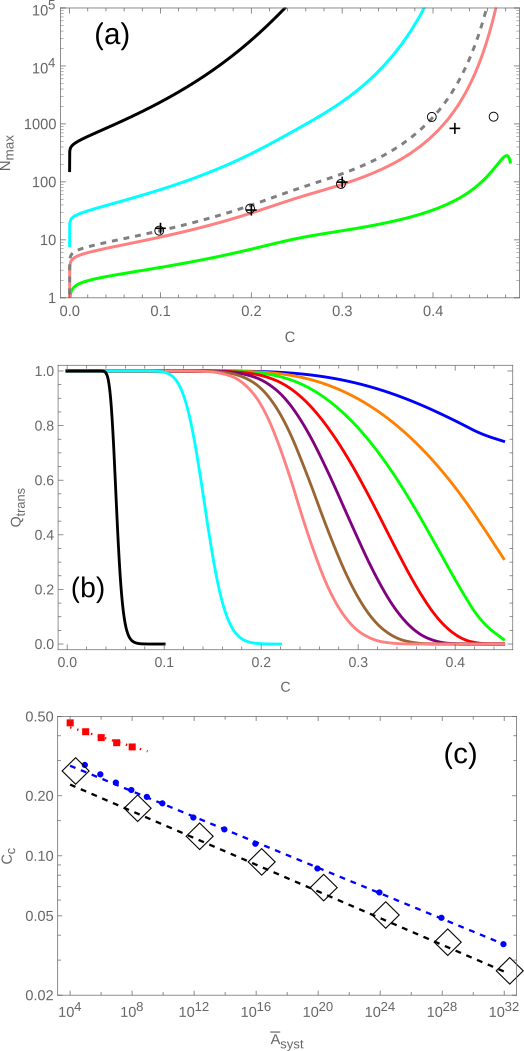}
	 	\caption{
   (a) Maximum cluster size. 
   The circles are obtained with the data of \cref{fig:cluster_size_distrib}, where
   $10^4$ simulations were performed in a system
   of size $\bar{A}_{syst}=10^5$. The pluses
   correspond to $10^5$ simulations with $\bar{A}_{syst}=10^4$.
   We therefore use $10^4\times 10^5=10^9\rightarrow \bar{A}_{syst}$
   in \cref{eq:cond_extreme_value}, leading the dashed line, in good agreement with simulation data, 
   except for the two last points at high $C$.  
   Solid lines from top to bottom correspond to \cref{eq:cond_extreme_value} with
   $\bar{A}_{syst}=10^4$, $10^8$, $10^{16}$, $10^{32}$. 
   (b) Probability $Q_{trans}$ as a function of $C$. From right to left (or top to bottom), we have $\bar{A}_{syst}=10^{2+n}$ with $n=0,..,6$, 
   and $10^{16}$, $10^{32}$.   
   (c) Transition threshold $C_c$ as a function of $\bar{A}_{syst}$. 
   Red squares: simulation results;
   Blue dots: analytical model;
   Black diamonds: solution of the equation
   $P_{av}=1/2$;
   Dashed lines are guides to the eye:  
    from top to bottom 
   $0.58 x^{-0.028}$, $0.38 x^{-0.032}$ 
   and $0.31 x^{-0.0335}$.
   }
	 	\label{fig:results_analytical_Cc}
	 \end{figure}

Then, using \cref{eq:cond_extreme_value}, \cref{eq:mu1_N0} and \cref{eq:N0_dist},
we can  directly evaluate $\mu_1$ and $N_{max}$.
The expression \cref{eq:cond_extreme_value} for $N_{max}$ 
is found to be in good agreement
with our numerical results, as seen from \cref{fig:results_analytical_Cc}(a).

However, the evaluation of 
$\tilde{N}_{max}=\bar{A}_{N_{max}}$ and $\tilde\mu_{n}$ requires
a careful study of the dependence of $A_N$ on $N$. 
Since the concentration of discs is homogeneous in the plane,
we expect  asymptotic behavior
$\bar{A}_{N\rightarrow \infty}\rightarrow N/C$.
However, as seen in \cref{fig:results_analytical}(b), 
this behavior is observed only for the largest clusters
and for $C\geq 0.4$. 
To account for the deviations at small $C$ and $N$, 
we propose a functional form
\begin{align}
\bar{A}_N= 1+ \frac{N-1}{C}(1-{\rm e}^{-a_0(N+a_3)^{a_1}C^{a_2}}).
\label{eq:ansatz_AN}
\end{align}
where $a_i$, $i=0,..,3$ are constants.
This expression presents the expected asymptotic behavior
for large $N$, and also imposes the correct behavior
for $N=1$, i.e., $\bar{A}_{N=1}=1$.
Moreover, we see that $\bar{A}_N$ does not diverge
 for $C\rightarrow 0$ when $a_2\geq 1$.  
The behavior of $\bar{A}_N$  is caught for all values of $C$
with $a_0=0.6$, $a_1=0.42$, $a_2=1.1$, and $a_3=1.3$, as seen in 
\cref{fig:results_analytical}(c). 

Using \cref{eq:ansatz_AN}, we can now determine 
\begin{align}
\bar{R}_{max}=\bar{A}_{N_{max}}^{1/2},
\end{align}
and
\begin{align}
\tilde{\mu}_k=\sum_{N=1}^\infty (C\bar{A}_N)^{k} P(N).
\label{eq:mu_nt_num}
\end{align}

\subsection{Results}

The evaluation of $Q_{trans(C)}$ leads to a behavior 
that shares similarities with the full numerical simulations
reported in the main text. Indeed, a transition is seen as a sharp drop
of $Q_{trans}$ in 
\cref{fig:results_analytical_Cc}(b).
When the system size $\bar{A}_{syst}$ is increased,
average transition value and transition width decrease
as reported in \cref{fig:transition_average}.

A limitation of the analytical model 
is the breakdown of the analytical description when 
$C\gtrsim 0.45$, due to the divergence 
of $\mu_1$ when $\delta\rightarrow 2$.
This effect is important for small $\bar{A}_{syst}$, when the transition
region spreads to large values of $C$.
However, this does not affect the transition
for large enough $\bar{A}_{syst}$.

The average location of the transition can be inferred
from the transition probability density $-dQ_{trans}/dC$,
leading to a transition at
\begin{align}
C_c=-\int_{0}^{C_{max}\rightarrow \infty} 
\hspace{-1.cm}dC\; C \frac{d}{dC} Q_{trans}
=\int_{0}^{C_{max}\rightarrow \infty} 
\hspace{-1.cm}dC\;  Q_{trans}
\end{align}
The last equality is obtained using integration 
by part in the limit $C_{max}\gg C_c$.
The numerical estimate of $C_c$ is not sensitive to the 
precise choice of $C_{max}$ as long as $C_{max}\gg C_c$.

The predicted value of $C_c$ is reported in
\cref{fig:results_analytical_Cc}(c).
We find that the predicted transition is quantitatively lower than that observed in direct simulations in the range $A_{syst}\leq 10^8$.
In addition, the decrease of $C_c$ with increasing $A_{syst}$
in simulations is slower than in the analytical model.

As an interesting result reported in \cref{fig:results_analytical_Cc}(c), 
the average value of the transition can be recovered by the condition $P_{av}=1/2$.
Hence, within our model, the transition is 
mainly controlled by the possibility for a macroscopic avalanche 
to spread through the whole system rather than 
being controlled by the details of the
initial growth dynamics of the avalanche (described
by $P_{start}$).
Note that $P_{start}$ is by far the most difficult 
quantity to evaluate.

\end{appendix}	

	\bibliography{biblio_article}

\begin{thebibliography}{52}%
\makeatletter
\providecommand \@ifxundefined [1]{%
 \@ifx{#1\undefined}
}%
\providecommand \@ifnum [1]{%
 \ifnum #1\expandafter \@firstoftwo
 \else \expandafter \@secondoftwo
 \fi
}%
\providecommand \@ifx [1]{%
 \ifx #1\expandafter \@firstoftwo
 \else \expandafter \@secondoftwo
 \fi
}%
\providecommand \natexlab [1]{#1}%
\providecommand \enquote  [1]{``#1''}%
\providecommand \bibnamefont  [1]{#1}%
\providecommand \bibfnamefont [1]{#1}%
\providecommand \citenamefont [1]{#1}%
\providecommand \href@noop [0]{\@secondoftwo}%
\providecommand \href [0]{\begingroup \@sanitize@url \@href}%
\providecommand \@href[1]{\@@startlink{#1}\@@href}%
\providecommand \@@href[1]{\endgroup#1\@@endlink}%
\providecommand \@sanitize@url [0]{\catcode `\\12\catcode `\$12\catcode
  `\&12\catcode `\#12\catcode `\^12\catcode `\_12\catcode `\%12\relax}%
\providecommand \@@startlink[1]{}%
\providecommand \@@endlink[0]{}%
\providecommand \url  [0]{\begingroup\@sanitize@url \@url }%
\providecommand \@url [1]{\endgroup\@href {#1}{\urlprefix }}%
\providecommand \urlprefix  [0]{URL }%
\providecommand \Eprint [0]{\href }%
\providecommand \doibase [0]{https://doi.org/}%
\providecommand \selectlanguage [0]{\@gobble}%
\providecommand \bibinfo  [0]{\@secondoftwo}%
\providecommand \bibfield  [0]{\@secondoftwo}%
\providecommand \translation [1]{[#1]}%
\providecommand \BibitemOpen [0]{}%
\providecommand \bibitemStop [0]{}%
\providecommand \bibitemNoStop [0]{.\EOS\space}%
\providecommand \EOS [0]{\spacefactor3000\relax}%
\providecommand \BibitemShut  [1]{\csname bibitem#1\endcsname}%
\let\auto@bib@innerbib\@empty
\bibitem [{\citenamefont {Livi}\ and\ \citenamefont {Politi}(2017)}]{Livi2017}%
  \BibitemOpen
  \bibfield  {author} {\bibinfo {author} {\bibfnamefont {R.}~\bibnamefont
  {Livi}}\ and\ \bibinfo {author} {\bibfnamefont {P.}~\bibnamefont {Politi}},\
  }\href {https://doi.org/10.1017/9781107278974} {\emph {\bibinfo {title}
  {Nonequilibrium Statistical Physics: A Modern Perspective}}}\ (\bibinfo
  {publisher} {Cambridge University Press},\ \bibinfo {year}
  {2017})\BibitemShut {NoStop}%
\bibitem [{\citenamefont {Krapivsky}\ \emph {et~al.}(2010)\citenamefont
  {Krapivsky}, \citenamefont {Redner},\ and\ \citenamefont
  {Ben-Naim}}]{Krapivsky2010}%
  \BibitemOpen
  \bibfield  {author} {\bibinfo {author} {\bibfnamefont {P.~L.}\ \bibnamefont
  {Krapivsky}}, \bibinfo {author} {\bibfnamefont {S.}~\bibnamefont {Redner}},\
  and\ \bibinfo {author} {\bibfnamefont {E.}~\bibnamefont {Ben-Naim}},\ }\href
  {https://doi.org/10.1017/CBO9780511780516} {\emph {\bibinfo {title} {A
  Kinetic View of Statistical Physics}}}\ (\bibinfo  {publisher} {Cambridge
  University Press},\ \bibinfo {year} {2010})\BibitemShut {NoStop}%
\bibitem [{\citenamefont {Aizenman}\ and\ \citenamefont
  {Lebowitz}(1988)}]{Aizenman1988}%
  \BibitemOpen
  \bibfield  {author} {\bibinfo {author} {\bibfnamefont {M.}~\bibnamefont
  {Aizenman}}\ and\ \bibinfo {author} {\bibfnamefont {J.~L.}\ \bibnamefont
  {Lebowitz}},\ }\bibfield  {title} {\bibinfo {title} {Metastability effects in
  bootstrap percolation},\ }\href {https://doi.org/10.1088/0305-4470/21/19/017}
  {\bibfield  {journal} {\bibinfo  {journal} {Journal of Physics A:
  Mathematical and General}\ }\textbf {\bibinfo {volume} {21}},\ \bibinfo
  {pages} {3801} (\bibinfo {year} {1988})}\BibitemShut {NoStop}%
\bibitem [{\citenamefont {Adler}(1991)}]{Adler1991}%
  \BibitemOpen
  \bibfield  {author} {\bibinfo {author} {\bibfnamefont {J.}~\bibnamefont
  {Adler}},\ }\bibfield  {title} {\bibinfo {title} {Bootstrap percolation},\
  }\href {https://doi.org/10.1016/0378-4371(91)90295-N} {\bibfield  {journal}
  {\bibinfo  {journal} {Physica A: Statistical Mechanics and its Applications}\
  }\textbf {\bibinfo {volume} {171}},\ \bibinfo {pages} {453} (\bibinfo {year}
  {1991})}\BibitemShut {NoStop}%
\bibitem [{\citenamefont {De~Gregorio}\ \emph {et~al.}(2021)\citenamefont
  {De~Gregorio}, \citenamefont {Lawlor},\ and\ \citenamefont
  {Dawson}}]{DeGregorio2021}%
  \BibitemOpen
  \bibfield  {author} {\bibinfo {author} {\bibfnamefont {P.}~\bibnamefont
  {De~Gregorio}}, \bibinfo {author} {\bibfnamefont {A.}~\bibnamefont
  {Lawlor}},\ and\ \bibinfo {author} {\bibfnamefont {K.~A.}\ \bibnamefont
  {Dawson}},\ }\bibinfo {title} {Bootstrap percolation},\ in\ \href
  {https://doi.org/10.1007/978-1-0716-1457-0_41} {\emph {\bibinfo {booktitle}
  {Complex Media and Percolation Theory}}},\ \bibinfo {editor} {edited by\
  \bibinfo {editor} {\bibfnamefont {M.}~\bibnamefont {Sahimi}}\ and\ \bibinfo
  {editor} {\bibfnamefont {A.~G.}\ \bibnamefont {Hunt}}}\ (\bibinfo
  {publisher} {Springer US},\ \bibinfo {address} {New York, NY},\ \bibinfo
  {year} {2021})\ pp.\ \bibinfo {pages} {149--173}\BibitemShut {NoStop}%
\bibitem [{\citenamefont {Holroyd}(2003)}]{Holroyd2003}%
  \BibitemOpen
  \bibfield  {author} {\bibinfo {author} {\bibfnamefont {A.~E.}\ \bibnamefont
  {Holroyd}},\ }\bibfield  {title} {\bibinfo {title} {{Sharp metastability
  threshold for two-dimensional bootstrap percolation}},\ }\href
  {https://doi.org/10.1007/s00440-002-0239-x} {\bibfield  {journal} {\bibinfo
  {journal} {Probability Theory and Related Fields}\ }\textbf {\bibinfo
  {volume} {125}},\ \bibinfo {pages} {195} (\bibinfo {year}
  {2003})}\BibitemShut {NoStop}%
\bibitem [{\citenamefont {De~Gregorio}\ \emph {et~al.}(2005)\citenamefont
  {De~Gregorio}, \citenamefont {Lawlor}, \citenamefont {Bradley},\ and\
  \citenamefont {Dawson}}]{DeGregorio2005}%
  \BibitemOpen
  \bibfield  {author} {\bibinfo {author} {\bibfnamefont {P.}~\bibnamefont
  {De~Gregorio}}, \bibinfo {author} {\bibfnamefont {A.}~\bibnamefont {Lawlor}},
  \bibinfo {author} {\bibfnamefont {P.}~\bibnamefont {Bradley}},\ and\ \bibinfo
  {author} {\bibfnamefont {K.~A.}\ \bibnamefont {Dawson}},\ }\bibfield  {title}
  {\bibinfo {title} {Exact solution of a jamming transition: Closed equations
  for a bootstrap percolation problem},\ }\href
  {https://doi.org/10.1073/pnas.0408756102} {\bibfield  {journal} {\bibinfo
  {journal} {Proceedings of the National Academy of Sciences}\ }\textbf
  {\bibinfo {volume} {102}},\ \bibinfo {pages} {5669} (\bibinfo {year}
  {2005})},\ \Eprint
  {https://arxiv.org/abs/https://www.pnas.org/content/102/16/5669.full.pdf}
  {https://www.pnas.org/content/102/16/5669.full.pdf} \BibitemShut {NoStop}%
\bibitem [{\citenamefont {Manna}(1998)}]{Manna1998}%
  \BibitemOpen
  \bibfield  {author} {\bibinfo {author} {\bibfnamefont {S.}~\bibnamefont
  {Manna}},\ }\bibfield  {title} {\bibinfo {title} {Abelian cascade dynamics in
  bootstrap percolation},\ }\href
  {https://doi.org/10.1016/S0378-4371(98)00346-X} {\bibfield  {journal}
  {\bibinfo  {journal} {Physica A: Statistical Mechanics and its Applications}\
  }\textbf {\bibinfo {volume} {261}},\ \bibinfo {pages} {351} (\bibinfo {year}
  {1998})}\BibitemShut {NoStop}%
\bibitem [{\citenamefont {Farrow}\ \emph {et~al.}(2005)\citenamefont {Farrow},
  \citenamefont {Duxbury},\ and\ \citenamefont {Moukarzel}}]{Farrow2005}%
  \BibitemOpen
  \bibfield  {author} {\bibinfo {author} {\bibfnamefont {C.}~\bibnamefont
  {Farrow}}, \bibinfo {author} {\bibfnamefont {P.~M.}\ \bibnamefont
  {Duxbury}},\ and\ \bibinfo {author} {\bibfnamefont {C.~F.}\ \bibnamefont
  {Moukarzel}},\ }\bibfield  {title} {\bibinfo {title} {Culling avalanches in
  bootstrap percolation},\ }\href {https://doi.org/10.1103/PhysRevE.72.066109}
  {\bibfield  {journal} {\bibinfo  {journal} {Phys. Rev. E}\ }\textbf {\bibinfo
  {volume} {72}},\ \bibinfo {pages} {066109} (\bibinfo {year}
  {2005})}\BibitemShut {NoStop}%
\bibitem [{\citenamefont {Farrow}\ \emph {et~al.}(2007)\citenamefont {Farrow},
  \citenamefont {Shukla},\ and\ \citenamefont {Duxbury}}]{Farrow2007}%
  \BibitemOpen
  \bibfield  {author} {\bibinfo {author} {\bibfnamefont {C.~L.}\ \bibnamefont
  {Farrow}}, \bibinfo {author} {\bibfnamefont {P.}~\bibnamefont {Shukla}},\
  and\ \bibinfo {author} {\bibfnamefont {P.~M.}\ \bibnamefont {Duxbury}},\
  }\bibfield  {title} {\bibinfo {title} {Dynamics of k-core percolation},\
  }\href {https://doi.org/10.1088/1751-8113/40/27/f02} {\bibfield  {journal}
  {\bibinfo  {journal} {Journal of Physics A: Mathematical and Theoretical}\
  }\textbf {\bibinfo {volume} {40}},\ \bibinfo {pages} {F581} (\bibinfo {year}
  {2007})}\BibitemShut {NoStop}%
\bibitem [{\citenamefont {D~Souza}\ and\ \citenamefont
  {Nagler}(2015)}]{DSouza2015}%
  \BibitemOpen
  \bibfield  {author} {\bibinfo {author} {\bibfnamefont {R.~M.}\ \bibnamefont
  {D~Souza}}\ and\ \bibinfo {author} {\bibfnamefont {J.}~\bibnamefont
  {Nagler}},\ }\bibfield  {title} {\bibinfo {title} {Anomalous critical and
  supercritical phenomena in explosive percolation},\ }\href
  {https://doi.org/10.1038/nphys3378} {\bibfield  {journal} {\bibinfo
  {journal} {Nature Physics}\ }\textbf {\bibinfo {volume} {11}},\ \bibinfo
  {pages} {531} (\bibinfo {year} {2015})}\BibitemShut {NoStop}%
\bibitem [{\citenamefont {D'Souza}\ \emph {et~al.}(2019)\citenamefont
  {D'Souza}, \citenamefont {G{\'o}mez-Gardenes}, \citenamefont {Nagler},\ and\
  \citenamefont {Arenas}}]{DSouza2019}%
  \BibitemOpen
  \bibfield  {author} {\bibinfo {author} {\bibfnamefont {R.~M.}\ \bibnamefont
  {D'Souza}}, \bibinfo {author} {\bibfnamefont {J.}~\bibnamefont
  {G{\'o}mez-Gardenes}}, \bibinfo {author} {\bibfnamefont {J.}~\bibnamefont
  {Nagler}},\ and\ \bibinfo {author} {\bibfnamefont {A.}~\bibnamefont
  {Arenas}},\ }\bibfield  {title} {\bibinfo {title} {Explosive phenomena in
  complex networks},\ }\href {https://doi.org/10.1080/00018732.2019.1650450}
  {\bibfield  {journal} {\bibinfo  {journal} {Advances in Physics}\ }\textbf
  {\bibinfo {volume} {68}},\ \bibinfo {pages} {123} (\bibinfo {year}
  {2019})}\BibitemShut {NoStop}%
\bibitem [{\citenamefont {Sethna}\ \emph {et~al.}(1993)\citenamefont {Sethna},
  \citenamefont {Dahmen}, \citenamefont {Kartha}, \citenamefont {Krumhansl},
  \citenamefont {Roberts},\ and\ \citenamefont {Shore}}]{Sethna1993}%
  \BibitemOpen
  \bibfield  {author} {\bibinfo {author} {\bibfnamefont {J.~P.}\ \bibnamefont
  {Sethna}}, \bibinfo {author} {\bibfnamefont {K.}~\bibnamefont {Dahmen}},
  \bibinfo {author} {\bibfnamefont {S.}~\bibnamefont {Kartha}}, \bibinfo
  {author} {\bibfnamefont {J.~A.}\ \bibnamefont {Krumhansl}}, \bibinfo {author}
  {\bibfnamefont {B.~W.}\ \bibnamefont {Roberts}},\ and\ \bibinfo {author}
  {\bibfnamefont {J.~D.}\ \bibnamefont {Shore}},\ }\bibfield  {title} {\bibinfo
  {title} {Hysteresis and hierarchies: Dynamics of disorder-driven first-order
  phase transformations},\ }\href {https://doi.org/10.1103/PhysRevLett.70.3347}
  {\bibfield  {journal} {\bibinfo  {journal} {Phys. Rev. Lett.}\ }\textbf
  {\bibinfo {volume} {70}},\ \bibinfo {pages} {3347} (\bibinfo {year}
  {1993})}\BibitemShut {NoStop}%
\bibitem [{\citenamefont {Rosinberg}\ \emph {et~al.}(2003)\citenamefont
  {Rosinberg}, \citenamefont {Kierlik},\ and\ \citenamefont
  {Tarjus}}]{Rosinberg2003}%
  \BibitemOpen
  \bibfield  {author} {\bibinfo {author} {\bibfnamefont {M.~L.}\ \bibnamefont
  {Rosinberg}}, \bibinfo {author} {\bibfnamefont {E.}~\bibnamefont {Kierlik}},\
  and\ \bibinfo {author} {\bibfnamefont {G.}~\bibnamefont {Tarjus}},\
  }\bibfield  {title} {\bibinfo {title} {Percolation, depinning, and avalanches
  in capillary condensation of gases in disordered porous solids},\ }\href
  {https://doi.org/10.1209/epl/i2003-00407-y} {\bibfield  {journal} {\bibinfo
  {journal} {Europhysics Letters ({EPL})}\ }\textbf {\bibinfo {volume} {62}},\
  \bibinfo {pages} {377} (\bibinfo {year} {2003})}\BibitemShut {NoStop}%
\bibitem [{\citenamefont {Nandi}\ \emph {et~al.}(2016)\citenamefont {Nandi},
  \citenamefont {Biroli},\ and\ \citenamefont {Tarjus}}]{Nandi2016}%
  \BibitemOpen
  \bibfield  {author} {\bibinfo {author} {\bibfnamefont {S.~K.}\ \bibnamefont
  {Nandi}}, \bibinfo {author} {\bibfnamefont {G.}~\bibnamefont {Biroli}},\ and\
  \bibinfo {author} {\bibfnamefont {G.}~\bibnamefont {Tarjus}},\ }\bibfield
  {title} {\bibinfo {title} {Spinodals with disorder: From avalanches in random
  magnets to glassy dynamics},\ }\href
  {https://doi.org/10.1103/PhysRevLett.116.145701} {\bibfield  {journal}
  {\bibinfo  {journal} {Phys. Rev. Lett.}\ }\textbf {\bibinfo {volume} {116}},\
  \bibinfo {pages} {145701} (\bibinfo {year} {2016})}\BibitemShut {NoStop}%
\bibitem [{\citenamefont {Saito}(1996)}]{Saito1996}%
  \BibitemOpen
  \bibfield  {author} {\bibinfo {author} {\bibfnamefont {Y.}~\bibnamefont
  {Saito}},\ }\href {https://doi.org/10.1142/3261} {\emph {\bibinfo {title}
  {Statistical Physics of Crystal Growth}}}\ (\bibinfo  {publisher} {World
  Scientific, Singapore},\ \bibinfo {year} {1996})\BibitemShut {NoStop}%
\bibitem [{\citenamefont {Paterson}\ \emph {et~al.}(1995)\citenamefont
  {Paterson}, \citenamefont {Fermigier}, \citenamefont {Jenffer},\ and\
  \citenamefont {Limat}}]{Paterson1995}%
  \BibitemOpen
  \bibfield  {author} {\bibinfo {author} {\bibfnamefont {A.}~\bibnamefont
  {Paterson}}, \bibinfo {author} {\bibfnamefont {M.}~\bibnamefont {Fermigier}},
  \bibinfo {author} {\bibfnamefont {P.}~\bibnamefont {Jenffer}},\ and\ \bibinfo
  {author} {\bibfnamefont {L.}~\bibnamefont {Limat}},\ }\bibfield  {title}
  {\bibinfo {title} {Wetting on heterogeneous surfaces: Experiments in an
  imperfect hele-shaw cell},\ }\href {https://doi.org/10.1103/PhysRevE.51.1291}
  {\bibfield  {journal} {\bibinfo  {journal} {Phys. Rev. E}\ }\textbf {\bibinfo
  {volume} {51}},\ \bibinfo {pages} {1291} (\bibinfo {year}
  {1995})}\BibitemShut {NoStop}%
\bibitem [{\citenamefont {Horv{\'a}th}\ \emph {et~al.}(1991)\citenamefont
  {Horv{\'a}th}, \citenamefont {Family},\ and\ \citenamefont
  {Vicsek}}]{Horvath1991}%
  \BibitemOpen
  \bibfield  {author} {\bibinfo {author} {\bibfnamefont {V.~K.}\ \bibnamefont
  {Horv{\'a}th}}, \bibinfo {author} {\bibfnamefont {F.}~\bibnamefont
  {Family}},\ and\ \bibinfo {author} {\bibfnamefont {T.}~\bibnamefont
  {Vicsek}},\ }\bibfield  {title} {\bibinfo {title} {Dynamic scaling of the
  interface in two-phase viscous flows in porous media},\ }\href
  {https://doi.org/10.1088/0305-4470/24/1/006} {\bibfield  {journal} {\bibinfo
  {journal} {Journal of Physics A: Mathematical and General}\ }\textbf
  {\bibinfo {volume} {24}},\ \bibinfo {pages} {L25} (\bibinfo {year}
  {1991})}\BibitemShut {NoStop}%
\bibitem [{\citenamefont {Rubio}\ \emph {et~al.}(1989)\citenamefont {Rubio},
  \citenamefont {Edwards}, \citenamefont {Dougherty},\ and\ \citenamefont
  {Gollub}}]{Rubio1989}%
  \BibitemOpen
  \bibfield  {author} {\bibinfo {author} {\bibfnamefont {M.}~\bibnamefont
  {Rubio}}, \bibinfo {author} {\bibfnamefont {C.}~\bibnamefont {Edwards}},
  \bibinfo {author} {\bibfnamefont {A.}~\bibnamefont {Dougherty}},\ and\
  \bibinfo {author} {\bibfnamefont {J.~P.}\ \bibnamefont {Gollub}},\ }\bibfield
   {title} {\bibinfo {title} {Self-affine fractal interfaces from immiscible
  displacement in porous media},\ }\href
  {https://doi.org/10.1103/PhysRevLett.63.1685} {\bibfield  {journal} {\bibinfo
   {journal} {Physical review letters}\ }\textbf {\bibinfo {volume} {63}},\
  \bibinfo {pages} {1685} (\bibinfo {year} {1989})}\BibitemShut {NoStop}%
\bibitem [{\citenamefont {Lii-Rosales}\ \emph {et~al.}(2020)\citenamefont
  {Lii-Rosales}, \citenamefont {Han}, \citenamefont {Julien}, \citenamefont
  {Pierre-Louis}, \citenamefont {Jing}, \citenamefont {Wan}, \citenamefont
  {Tringides}, \citenamefont {Evans},\ and\ \citenamefont
  {Thiel}}]{LiiRosales2020}%
  \BibitemOpen
  \bibfield  {author} {\bibinfo {author} {\bibfnamefont {A.}~\bibnamefont
  {Lii-Rosales}}, \bibinfo {author} {\bibfnamefont {Y.}~\bibnamefont {Han}},
  \bibinfo {author} {\bibfnamefont {S.~E.}\ \bibnamefont {Julien}}, \bibinfo
  {author} {\bibfnamefont {O.}~\bibnamefont {Pierre-Louis}}, \bibinfo {author}
  {\bibfnamefont {D.}~\bibnamefont {Jing}}, \bibinfo {author} {\bibfnamefont
  {K.-T.}\ \bibnamefont {Wan}}, \bibinfo {author} {\bibfnamefont {M.~C.}\
  \bibnamefont {Tringides}}, \bibinfo {author} {\bibfnamefont {J.~W.}\
  \bibnamefont {Evans}},\ and\ \bibinfo {author} {\bibfnamefont {P.~A.}\
  \bibnamefont {Thiel}},\ }\bibfield  {title} {\bibinfo {title} {Shapes of fe
  nanocrystals encapsulated at the graphite surface},\ }\href
  {https://doi.org/10.1088/1367-2630/ab687a} {\bibfield  {journal} {\bibinfo
  {journal} {New Journal of Physics}\ }\textbf {\bibinfo {volume} {22}},\
  \bibinfo {pages} {023016} (\bibinfo {year} {2020})}\BibitemShut {NoStop}%
\bibitem [{\citenamefont {Zong}\ \emph {et~al.}(2010)\citenamefont {Zong},
  \citenamefont {Chen}, \citenamefont {Dokmeci},\ and\ \citenamefont
  {Wan}}]{Zong2010}%
  \BibitemOpen
  \bibfield  {author} {\bibinfo {author} {\bibfnamefont {Z.}~\bibnamefont
  {Zong}}, \bibinfo {author} {\bibfnamefont {C.-L.}\ \bibnamefont {Chen}},
  \bibinfo {author} {\bibfnamefont {M.~R.}\ \bibnamefont {Dokmeci}},\ and\
  \bibinfo {author} {\bibfnamefont {K.-t.}\ \bibnamefont {Wan}},\ }\bibfield
  {title} {\bibinfo {title} {{Direct measurement of graphene adhesion on
  silicon surface by intercalation of nanoparticles}},\ }\bibfield  {journal}
  {\bibinfo  {journal} {Journal of Applied Physics}\ }\textbf {\bibinfo
  {volume} {107}},\ \href {https://doi.org/10.1063/1.3294960}
  {10.1063/1.3294960} (\bibinfo {year} {2010}),\ \bibinfo {note} {026104},\
  \Eprint
  {https://arxiv.org/abs/https://pubs.aip.org/aip/jap/article-pdf/doi/10.1063/1.3294960/15044015/026104\_1\_online.pdf}
  {https://pubs.aip.org/aip/jap/article-pdf/doi/10.1063/1.3294960/15044015/026104\_1\_online.pdf}
  \BibitemShut {NoStop}%
\bibitem [{\citenamefont {Wang}\ \emph {et~al.}(2016)\citenamefont {Wang},
  \citenamefont {Sorescu}, \citenamefont {Jeon}, \citenamefont {Belianinov},
  \citenamefont {Kalinin}, \citenamefont {Baddorf},\ and\ \citenamefont
  {Maksymovych}}]{Wang2016}%
  \BibitemOpen
  \bibfield  {author} {\bibinfo {author} {\bibfnamefont {J.}~\bibnamefont
  {Wang}}, \bibinfo {author} {\bibfnamefont {D.~C.}\ \bibnamefont {Sorescu}},
  \bibinfo {author} {\bibfnamefont {S.}~\bibnamefont {Jeon}}, \bibinfo {author}
  {\bibfnamefont {A.}~\bibnamefont {Belianinov}}, \bibinfo {author}
  {\bibfnamefont {S.~V.}\ \bibnamefont {Kalinin}}, \bibinfo {author}
  {\bibfnamefont {A.~P.}\ \bibnamefont {Baddorf}},\ and\ \bibinfo {author}
  {\bibfnamefont {P.}~\bibnamefont {Maksymovych}},\ }\bibfield  {title}
  {\bibinfo {title} {Atomic intercalation to measure adhesion of graphene on
  graphite},\ }\href {https://doi.org/10.1038/ncomms13263} {\bibfield
  {journal} {\bibinfo  {journal} {Nature communications}\ }\textbf {\bibinfo
  {volume} {7}},\ \bibinfo {pages} {1} (\bibinfo {year} {2016})}\BibitemShut
  {NoStop}%
\bibitem [{\citenamefont {Yamamoto}\ \emph {et~al.}(2012)\citenamefont
  {Yamamoto}, \citenamefont {Pierre-Louis}, \citenamefont {Huang},
  \citenamefont {Fuhrer}, \citenamefont {Einstein},\ and\ \citenamefont
  {Cullen}}]{Yamamoto2012}%
  \BibitemOpen
  \bibfield  {author} {\bibinfo {author} {\bibfnamefont {M.}~\bibnamefont
  {Yamamoto}}, \bibinfo {author} {\bibfnamefont {O.}~\bibnamefont
  {Pierre-Louis}}, \bibinfo {author} {\bibfnamefont {J.}~\bibnamefont {Huang}},
  \bibinfo {author} {\bibfnamefont {M.~S.}\ \bibnamefont {Fuhrer}}, \bibinfo
  {author} {\bibfnamefont {T.~L.}\ \bibnamefont {Einstein}},\ and\ \bibinfo
  {author} {\bibfnamefont {W.~G.}\ \bibnamefont {Cullen}},\ }\bibfield  {title}
  {\bibinfo {title} {{The princess and the pea at the nanoscale: Wrinkling and
  delamination of graphene on nanoparticles}},\ }\bibfield  {journal} {\bibinfo
   {journal} {Physical Review X}\ }\textbf {\bibinfo {volume} {2}},\ \href
  {https://doi.org/10.1103/PhysRevX.2.041018} {10.1103/PhysRevX.2.041018}
  (\bibinfo {year} {2012}),\ \Eprint {https://arxiv.org/abs/1201.5667}
  {arXiv:1201.5667} \BibitemShut {NoStop}%
\bibitem [{\citenamefont {Guedda}\ \emph {et~al.}(2016)\citenamefont {Guedda},
  \citenamefont {Alaa},\ and\ \citenamefont {Benlahsen}}]{Guedda2016}%
  \BibitemOpen
  \bibfield  {author} {\bibinfo {author} {\bibfnamefont {M.}~\bibnamefont
  {Guedda}}, \bibinfo {author} {\bibfnamefont {N.}~\bibnamefont {Alaa}},\ and\
  \bibinfo {author} {\bibfnamefont {M.}~\bibnamefont {Benlahsen}},\ }\bibfield
  {title} {\bibinfo {title} {Analytical results for the wrinkling of graphene
  on nanoparticles},\ }\href {https://doi.org/10.1103/PhysRevE.94.042806}
  {\bibfield  {journal} {\bibinfo  {journal} {Physical Review E}\ }\textbf
  {\bibinfo {volume} {94}},\ \bibinfo {pages} {042806} (\bibinfo {year}
  {2016})}\BibitemShut {NoStop}%
\bibitem [{\citenamefont {Zhang}\ and\ \citenamefont
  {Arroyo}(2014)}]{Zhang2014}%
  \BibitemOpen
  \bibfield  {author} {\bibinfo {author} {\bibfnamefont {K.}~\bibnamefont
  {Zhang}}\ and\ \bibinfo {author} {\bibfnamefont {M.}~\bibnamefont {Arroyo}},\
  }\bibfield  {title} {\bibinfo {title} {Understanding and strain-engineering
  wrinkle networks in supported graphene through simulations},\ }\href
  {https://doi.org/10.1016/j.jmps.2014.07.012} {\bibfield  {journal} {\bibinfo
  {journal} {Journal of the Mechanics and Physics of Solids}\ }\textbf
  {\bibinfo {volume} {72}},\ \bibinfo {pages} {61} (\bibinfo {year}
  {2014})}\BibitemShut {NoStop}%
\bibitem [{\citenamefont {Pereira}\ \emph {et~al.}(2010)\citenamefont
  {Pereira}, \citenamefont {Castro~Neto}, \citenamefont {Liang},\ and\
  \citenamefont {Mahadevan}}]{Pereira2010}%
  \BibitemOpen
  \bibfield  {author} {\bibinfo {author} {\bibfnamefont {V.~M.}\ \bibnamefont
  {Pereira}}, \bibinfo {author} {\bibfnamefont {A.~H.}\ \bibnamefont
  {Castro~Neto}}, \bibinfo {author} {\bibfnamefont {H.~Y.}\ \bibnamefont
  {Liang}},\ and\ \bibinfo {author} {\bibfnamefont {L.}~\bibnamefont
  {Mahadevan}},\ }\bibfield  {title} {\bibinfo {title} {Geometry, mechanics,
  and electronics of singular structures and wrinkles in graphene},\ }\href
  {https://doi.org/10.1103/PhysRevLett.105.156603} {\bibfield  {journal}
  {\bibinfo  {journal} {Phys. Rev. Lett.}\ }\textbf {\bibinfo {volume} {105}},\
  \bibinfo {pages} {156603} (\bibinfo {year} {2010})}\BibitemShut {NoStop}%
\bibitem [{\citenamefont {Mollison}(1972)}]{Mollison1972}%
  \BibitemOpen
  \bibfield  {author} {\bibinfo {author} {\bibfnamefont {D.}~\bibnamefont
  {Mollison}},\ }\bibfield  {title} {\bibinfo {title} {The rate of spatial
  propagation of simple epidemics},\ }in\ \href
  {https://digitalassets.lib.berkeley.edu/math/ucb/text/math_s6_v3_article-31.pdf}
  {\emph {\bibinfo {booktitle} {Proceedings of the Sixth Berkeley Symposium on
  Mathematical Statistics and Probability, Volume 3: Probability Theory}}}\
  (\bibinfo {organization} {University of California Press},\ \bibinfo {year}
  {1972})\ pp.\ \bibinfo {pages} {579--614}\BibitemShut {NoStop}%
\bibitem [{\citenamefont {Kuperman}\ and\ \citenamefont
  {Wio}(1999)}]{Kuperman1999}%
  \BibitemOpen
  \bibfield  {author} {\bibinfo {author} {\bibfnamefont {M.}~\bibnamefont
  {Kuperman}}\ and\ \bibinfo {author} {\bibfnamefont {H.}~\bibnamefont {Wio}},\
  }\bibfield  {title} {\bibinfo {title} {Front propagation in epidemiological
  models with spatial dependence},\ }\href
  {https://doi.org/https://doi.org/10.1016/S0378-4371(99)00284-8} {\bibfield
  {journal} {\bibinfo  {journal} {Physica A: Statistical Mechanics and its
  Applications}\ }\textbf {\bibinfo {volume} {272}},\ \bibinfo {pages} {206}
  (\bibinfo {year} {1999})}\BibitemShut {NoStop}%
\bibitem [{\citenamefont {Abramson}\ \emph {et~al.}(2003)\citenamefont
  {Abramson}, \citenamefont {Kenkre}, \citenamefont {Yates},\ and\
  \citenamefont {Parmenter}}]{Abramson2003}%
  \BibitemOpen
  \bibfield  {author} {\bibinfo {author} {\bibfnamefont {G.}~\bibnamefont
  {Abramson}}, \bibinfo {author} {\bibfnamefont {V.}~\bibnamefont {Kenkre}},
  \bibinfo {author} {\bibfnamefont {T.}~\bibnamefont {Yates}},\ and\ \bibinfo
  {author} {\bibfnamefont {R.}~\bibnamefont {Parmenter}},\ }\bibfield  {title}
  {\bibinfo {title} {Traveling waves of infection in the hantavirus
  epidemics},\ }\href
  {https://link.springer.com/article/10.1016/S0092-8240(03)00013-2} {\bibfield
  {journal} {\bibinfo  {journal} {Bulletin of mathematical biology}\ }\textbf
  {\bibinfo {volume} {65}},\ \bibinfo {pages} {519} (\bibinfo {year}
  {2003})}\BibitemShut {NoStop}%
\bibitem [{\citenamefont {Hanert}\ \emph {et~al.}(2011)\citenamefont {Hanert},
  \citenamefont {Schumacher},\ and\ \citenamefont
  {Deleersnijder}}]{Hanert2011}%
  \BibitemOpen
  \bibfield  {author} {\bibinfo {author} {\bibfnamefont {E.}~\bibnamefont
  {Hanert}}, \bibinfo {author} {\bibfnamefont {E.}~\bibnamefont {Schumacher}},\
  and\ \bibinfo {author} {\bibfnamefont {E.}~\bibnamefont {Deleersnijder}},\
  }\bibfield  {title} {\bibinfo {title} {Front dynamics in fractional-order
  epidemic models},\ }\href
  {https://doi.org/https://doi.org/10.1016/j.jtbi.2011.03.012} {\bibfield
  {journal} {\bibinfo  {journal} {Journal of Theoretical Biology}\ }\textbf
  {\bibinfo {volume} {279}},\ \bibinfo {pages} {9} (\bibinfo {year}
  {2011})}\BibitemShut {NoStop}%
\bibitem [{\citenamefont {Ben-Israel}\ and\ \citenamefont
  {Levin}(2006)}]{BenIsrael2006}%
  \BibitemOpen
  \bibfield  {author} {\bibinfo {author} {\bibfnamefont {A.}~\bibnamefont
  {Ben-Israel}}\ and\ \bibinfo {author} {\bibfnamefont {Y.}~\bibnamefont
  {Levin}},\ }\bibfield  {title} {\bibinfo {title} {The geometry of linear
  separability in data sets},\ }\href
  {https://doi.org/10.1016/j.laa.2005.08.014} {\bibfield  {journal} {\bibinfo
  {journal} {Linear algebra and its applications}\ }\textbf {\bibinfo {volume}
  {416}},\ \bibinfo {pages} {75} (\bibinfo {year} {2006})}\BibitemShut
  {NoStop}%
\bibitem [{\citenamefont {Hardy}(1996)}]{Hardy1996}%
  \BibitemOpen
  \bibfield  {author} {\bibinfo {author} {\bibfnamefont {A.}~\bibnamefont
  {Hardy}},\ }\bibfield  {title} {\bibinfo {title} {On the number of
  clusters},\ }\href {https://doi.org/10.1016/S0167-9473(96)00022-9} {\bibfield
   {journal} {\bibinfo  {journal} {Computational Statistics \& Data Analysis}\
  }\textbf {\bibinfo {volume} {23}},\ \bibinfo {pages} {83} (\bibinfo {year}
  {1996})}\BibitemShut {NoStop}%
\bibitem [{\citenamefont {Jos{\'e}-Garc{\'\i}a}\ and\ \citenamefont
  {G{\'o}mez-Flores}(2016)}]{JoseGarcia2016}%
  \BibitemOpen
  \bibfield  {author} {\bibinfo {author} {\bibfnamefont {A.}~\bibnamefont
  {Jos{\'e}-Garc{\'\i}a}}\ and\ \bibinfo {author} {\bibfnamefont
  {W.}~\bibnamefont {G{\'o}mez-Flores}},\ }\bibfield  {title} {\bibinfo {title}
  {Automatic clustering using nature-inspired metaheuristics: A survey},\
  }\href {https://doi.org/10.1016/j.asoc.2015.12.001} {\bibfield  {journal}
  {\bibinfo  {journal} {Applied Soft Computing}\ }\textbf {\bibinfo {volume}
  {41}},\ \bibinfo {pages} {192} (\bibinfo {year} {2016})}\BibitemShut
  {NoStop}%
\bibitem [{\citenamefont {Elizondo}(2006)}]{Elizondo2006}%
  \BibitemOpen
  \bibfield  {author} {\bibinfo {author} {\bibfnamefont {D.}~\bibnamefont
  {Elizondo}},\ }\bibfield  {title} {\bibinfo {title} {The linear separability
  problem: some testing methods},\ }\href
  {https://doi.org/10.1109/TNN.2005.860871} {\bibfield  {journal} {\bibinfo
  {journal} {IEEE Transactions on Neural Networks}\ }\textbf {\bibinfo {volume}
  {17}},\ \bibinfo {pages} {330} (\bibinfo {year} {2006})}\BibitemShut
  {NoStop}%
\bibitem [{\citenamefont {Steinwart}\ and\ \citenamefont
  {Christmann}(2008)}]{Steinwart2008}%
  \BibitemOpen
  \bibfield  {author} {\bibinfo {author} {\bibfnamefont {I.}~\bibnamefont
  {Steinwart}}\ and\ \bibinfo {author} {\bibfnamefont {A.}~\bibnamefont
  {Christmann}},\ }\href {https://doi.org/10.1007/978-0-387-77242-4} {\emph
  {\bibinfo {title} {Support vector machines}}}\ (\bibinfo  {publisher}
  {Springer Science \& Business Media},\ \bibinfo {year} {2008})\BibitemShut
  {NoStop}%
\bibitem [{\citenamefont {Wang}(2005)}]{Wang2005}%
  \BibitemOpen
  \bibfield  {author} {\bibinfo {author} {\bibfnamefont {L.}~\bibnamefont
  {Wang}},\ }\href {https://doi.org/10.1007/b95439} {\emph {\bibinfo {title}
  {Support vector machines: theory and applications}}},\ Vol.\ \bibinfo
  {volume} {177}\ (\bibinfo  {publisher} {Springer Science \& Business Media},\
  \bibinfo {year} {2005})\BibitemShut {NoStop}%
\bibitem [{\citenamefont {Jayaram}\ and\ \citenamefont
  {Fleyeh}(2016)}]{Jayaram2016}%
  \BibitemOpen
  \bibfield  {author} {\bibinfo {author} {\bibfnamefont {M.}~\bibnamefont
  {Jayaram}}\ and\ \bibinfo {author} {\bibfnamefont {H.}~\bibnamefont
  {Fleyeh}},\ }\bibfield  {title} {\bibinfo {title} {Convex hulls in image
  processing: a scoping review},\ }\href
  {https://www.diva-portal.org/smash/record.jsf?pid=diva2:931027} {\bibfield
  {journal} {\bibinfo  {journal} {American Journal of Intelligent Systems}\
  }\textbf {\bibinfo {volume} {6}},\ \bibinfo {pages} {48} (\bibinfo {year}
  {2016})}\BibitemShut {NoStop}%
\bibitem [{\citenamefont {Malladi}\ and\ \citenamefont
  {Sethian}(1995)}]{Malladi1995}%
  \BibitemOpen
  \bibfield  {author} {\bibinfo {author} {\bibfnamefont {R.}~\bibnamefont
  {Malladi}}\ and\ \bibinfo {author} {\bibfnamefont {J.~A.}\ \bibnamefont
  {Sethian}},\ }\bibfield  {title} {\bibinfo {title} {Image processing via
  level set curvature flow},\ }\href {https://doi.org/10.1073/pnas.92.15.7046}
  {\bibfield  {journal} {\bibinfo  {journal} {proceedings of the National
  Academy of sciences}\ }\textbf {\bibinfo {volume} {92}},\ \bibinfo {pages}
  {7046} (\bibinfo {year} {1995})}\BibitemShut {NoStop}%
\bibitem [{\citenamefont {Malladi}\ and\ \citenamefont
  {Sethian}(1996)}]{Malladi1996}%
  \BibitemOpen
  \bibfield  {author} {\bibinfo {author} {\bibfnamefont {R.}~\bibnamefont
  {Malladi}}\ and\ \bibinfo {author} {\bibfnamefont {J.~A.}\ \bibnamefont
  {Sethian}},\ }\bibfield  {title} {\bibinfo {title} {Image processing: Flows
  under min/max curvature and mean curvature},\ }\href
  {https://doi.org/10.1006/gmip.1996.0011} {\bibfield  {journal} {\bibinfo
  {journal} {Graphical models and image processing}\ }\textbf {\bibinfo
  {volume} {58}},\ \bibinfo {pages} {127} (\bibinfo {year} {1996})}\BibitemShut
  {NoStop}%
\bibitem [{Note1()}]{Note1}%
  \BibitemOpen
  \bibinfo {note} {More precisely, the centers of the discs are chosen with a
  continuous uniform probability distribution on the system area.}\BibitemShut
  {Stop}%
\bibitem [{\citenamefont {Gawlinski}\ and\ \citenamefont
  {Stanley}(1981)}]{Gawlinski1981}%
  \BibitemOpen
  \bibfield  {author} {\bibinfo {author} {\bibfnamefont {E.~T.}\ \bibnamefont
  {Gawlinski}}\ and\ \bibinfo {author} {\bibfnamefont {H.~E.}\ \bibnamefont
  {Stanley}},\ }\bibfield  {title} {\bibinfo {title} {{Continuum percolation in
  two dimensions: Monte Carlo tests of scaling and universality for
  non-interacting discs}},\ }\href {https://doi.org/10.1088/0305-4470/14/8/007}
  {\bibfield  {journal} {\bibinfo  {journal} {Journal of Physics A:
  Mathematical and General}\ }\textbf {\bibinfo {volume} {14}},\ \bibinfo
  {pages} {L291} (\bibinfo {year} {1981})}\BibitemShut {NoStop}%
\bibitem [{Note2()}]{Note2}%
  \BibitemOpen
  \bibinfo {note} {This relation is exact for periodic systems and
  asymptotically true in finite systems of large size when the effect of
  boundary conditions can be neglected.}\BibitemShut {Stop}%
\bibitem [{\citenamefont {Quintanilla}\ \emph {et~al.}(2000)\citenamefont
  {Quintanilla}, \citenamefont {Torquato},\ and\ \citenamefont
  {Ziff}}]{Quintanilla2000}%
  \BibitemOpen
  \bibfield  {author} {\bibinfo {author} {\bibfnamefont {J.}~\bibnamefont
  {Quintanilla}}, \bibinfo {author} {\bibfnamefont {S.}~\bibnamefont
  {Torquato}},\ and\ \bibinfo {author} {\bibfnamefont {R.~M.}\ \bibnamefont
  {Ziff}},\ }\bibfield  {title} {\bibinfo {title} {Efficient measurement of the
  percolation threshold for fully penetrable discs},\ }\href
  {https://doi.org/10.1088/0305-4470/33/42/104} {\bibfield  {journal} {\bibinfo
   {journal} {Journal of Physics A: Mathematical and General}\ }\textbf
  {\bibinfo {volume} {33}},\ \bibinfo {pages} {L399} (\bibinfo {year}
  {2000})}\BibitemShut {NoStop}%
\bibitem [{Note3()}]{Note3}%
  \BibitemOpen
  \bibinfo {note} {In circular systems, the centers of the discs are placed
  randomly in the system. Hence, when the center of a disc is close to the edge
  of the system, part of this discs may lie outside the system.}\BibitemShut
  {Stop}%
\bibitem [{\citenamefont {Stauffer}\ and\ \citenamefont
  {Aharony}(2018)}]{Stauffer2018}%
  \BibitemOpen
  \bibfield  {author} {\bibinfo {author} {\bibfnamefont {D.}~\bibnamefont
  {Stauffer}}\ and\ \bibinfo {author} {\bibfnamefont {A.}~\bibnamefont
  {Aharony}},\ }\href {https://doi.org/10.1201/9781315274386} {\emph {\bibinfo
  {title} {Introduction to percolation theory}}}\ (\bibinfo  {publisher}
  {Taylor \& Francis},\ \bibinfo {year} {2018})\BibitemShut {NoStop}%
\bibitem [{Note4()}]{Note4}%
  \BibitemOpen
  \bibinfo {note} {For each system size $\protect \bar {A}_{syst} \in
  \{10^4,10^5,10^6,10^7,10^8\}$, we run $\{10^5,10^4,10^3,10^3,60\}$
  realizations of our simulation. Then we extract the data for $C \in
  \{0.1,0.2,0.3,0.4,C_c=0.47\}$ for $\protect \bar {A}_{syst}=10^4$ and $C \in
  \{0.1,0.2,0.3,C_c\}$ for $\protect \bar {A}_{syst} \in
  \{10^5,10^6,10^7,10^8\}$, with $C_c \in \{0.423,0.394,0.372,0.355\}$. We
  exclude realizations where the system has already undergone a transition,
  corresponding to a significant number of runs mainly at $C=C_c$, where
  roughly half the realizations have transited.}\BibitemShut {Stop}%
\bibitem [{\citenamefont {R}\ and\ \citenamefont {C}(1984)}]{Zarcone1984}%
  \BibitemOpen
  \bibfield  {author} {\bibinfo {author} {\bibfnamefont {L.}~\bibnamefont {R}}\
  and\ \bibinfo {author} {\bibfnamefont {Z.}~\bibnamefont {C}},\ }\bibinfo
  {title} {Growth of clusters during imbibition in a network of capillaries},\
  in\ \href@noop {} {\emph {\bibinfo {booktitle} {Kinetics of Aggregation and
  Gelation}}},\ \bibinfo {editor} {edited by\ \bibinfo {editor} {\bibfnamefont
  {F.}~\bibnamefont {Family}}\ and\ \bibinfo {editor} {\bibfnamefont {D.~P.}\
  \bibnamefont {Landau}}}\ (\bibinfo  {publisher} {Elsevier},\ \bibinfo
  {address} {Amsterdam},\ \bibinfo {year} {1984})\ p.\ \bibinfo {pages}
  {177}\BibitemShut {NoStop}%
\bibitem [{\citenamefont {Adler}\ \emph {et~al.}(1989)\citenamefont {Adler},
  \citenamefont {Stauffer},\ and\ \citenamefont {Aharony}}]{Adler1989}%
  \BibitemOpen
  \bibfield  {author} {\bibinfo {author} {\bibfnamefont {J.}~\bibnamefont
  {Adler}}, \bibinfo {author} {\bibfnamefont {D.}~\bibnamefont {Stauffer}},\
  and\ \bibinfo {author} {\bibfnamefont {A.}~\bibnamefont {Aharony}},\
  }\bibfield  {title} {\bibinfo {title} {Comparison of bootstrap percolation
  models},\ }\href@noop {} {\bibfield  {journal} {\bibinfo  {journal} {Journal
  of Physics A: Mathematical and General}\ }\textbf {\bibinfo {volume} {22}},\
  \bibinfo {pages} {L297} (\bibinfo {year} {1989})}\BibitemShut {NoStop}%
\bibitem [{\citenamefont {Schwerin}(1929)}]{Schwerin1929}%
  \BibitemOpen
  \bibfield  {author} {\bibinfo {author} {\bibfnamefont {E.}~\bibnamefont
  {Schwerin}},\ }\bibfield  {title} {\bibinfo {title} {\"uber spannungen und
  form\"anderungen kreisringf\"ormiger membranen},\ }\href
  {https://doi.org/10.1002/zamm.19290090609} {\bibfield  {journal} {\bibinfo
  {journal} {Zeitschrift Angewandte Mathematik und Mechanik}\ }\textbf
  {\bibinfo {volume} {9}},\ \bibinfo {pages} {482} (\bibinfo {year}
  {1929})}\BibitemShut {NoStop}%
\bibitem [{\citenamefont {Komaragiri}\ \emph {et~al.}(2005)\citenamefont
  {Komaragiri}, \citenamefont {Begley},\ and\ \citenamefont
  {Simmonds}}]{Komaragiri2005}%
  \BibitemOpen
  \bibfield  {author} {\bibinfo {author} {\bibfnamefont {U.}~\bibnamefont
  {Komaragiri}}, \bibinfo {author} {\bibfnamefont {M.}~\bibnamefont {Begley}},\
  and\ \bibinfo {author} {\bibfnamefont {J.}~\bibnamefont {Simmonds}},\
  }\bibfield  {title} {\bibinfo {title} {The mechanical response of
  freestanding circular elastic films under point and pressure loads},\ }\href
  {https://doi.org/10.1115/1.1827246} {\bibfield  {journal} {\bibinfo
  {journal} {J. Appl. Mech.}\ }\textbf {\bibinfo {volume} {72}},\ \bibinfo
  {pages} {203} (\bibinfo {year} {2005})}\BibitemShut {NoStop}%
\bibitem [{Note5()}]{Note5}%
  \BibitemOpen
  \bibinfo {note} {The Hurwitz Zeta function is defined as $$ \zeta
  (s,a)=\DOTSB \sum@ \slimits@ _{n=0}^\infty \protect \frac {1}{(n+a)^s}.
  $$}\BibitemShut {NoStop}%
\bibitem [{Note6()}]{Note6}%
  \BibitemOpen
  \bibinfo {note} {Note however that we neglect the correlations betwen the
  clusters other than the growing cluster in all cases.}\BibitemShut {Stop}%
\end{thebibliography}%

\end{document}